\documentclass[twocolumn]{aastex631}

\usepackage{xcolor}

\shorttitle{LMC Bar Dynamics}
\shortauthors{Rathore et al.}

\usepackage{csquotes}
\usepackage{amsmath}

\newcommand{\vect}[1]{\boldsymbol{#1}}

\begin{document}
\title{Response of the LMC's Bar to a Recent SMC Collision and Implications for the SMC's Dark Matter Profile}

\correspondingauthor{Himansh Rathore}
\email{himansh@arizona.edu}

\author[0009-0009-0158-585X]{Himansh Rathore}
\affiliation{Department of Astronomy and Steward Observatory, University of Arizona, 933 North Cherry Avenue, Tucson, AZ 85721, USA}

\author[0000-0003-0715-2173]{Gurtina Besla}
\affiliation{Department of Astronomy and Steward Observatory, University of Arizona, 933 North Cherry Avenue, Tucson, AZ 85721, USA}

\author[0000-0003-2594-8052]{Kathryne J. Daniel}
\affiliation{Department of Astronomy and Steward Observatory, University of Arizona, 933 North Cherry Avenue, Tucson, AZ 85721, USA}

\author[0000-0002-0740-1507]{Leandro {Beraldo e Silva}}
\affiliation{Department of Astronomy and Steward Observatory, University of Arizona, 933 North Cherry Avenue, Tucson, AZ 85721, USA}
\affiliation{Observat{\'o}rio Nacional, Rio de Janeiro - RJ, 20921-400, Brasil}

\begin{abstract}
The LMC's stellar bar is offset from the outer disk center, tilted from the disk plane, and does not drive gas inflows. These properties are atypical of bars in gas-rich galaxies, yet the LMC bar's strength and radius are similar to typical barred galaxies. Using N-body hydrodynamic simulations, we show that the LMC's unusual bar is explainable if there was a recent collision (impact parameter $\approx$2 kpc) between the LMC and SMC. Pre-collision, the simulated bar is centered and co-planar. Post-collision, the simulated bar is offset ($\approx$1.5 kpc) and tilted ($\approx8.6^\circ$). The simulated bar offset reduces with time, and comparing with the observed offset ($\approx0.8$~kpc) suggests the timing of the true collision to be 150-200~Myr ago. 150 Myr post-collision, the LMC's bar is centered with its dark matter halo, whereas the outer disk center is separated from the dark matter center by $\approx1$~kpc. The SMC collision produces a tilted-ring structure for the simulated LMC, consistent with observations. Post-collision, the simulated LMC bar's pattern speed decreases by a factor of two. We also provide a generalizable framework to quantitatively compare the LMC's central gas distribution in different LMC-SMC interaction scenarios. We demonstrate that the SMC's torques on the LMC's bar during the collision are sufficient to explain the observed bar tilt, provided the SMC's total mass within 2~kpc was $(0.8-2.4) \times 10^9$ M$_\odot$. Therefore, the LMC bar's tilt constrains the SMC's pre-collision dark matter profile, and requires the SMC to be a dark matter-dominated galaxy.  
\end{abstract}

\keywords{\href{http://astrothesaurus.org/uat/903}{LMC (903)}; \href{http://astrothesaurus.org/uat/1468}{SMC (1468)}; \href{http://astrothesaurus.org/uat/2364}{Galaxy bars (2364)};  \href{http://astrothesaurus.org/uat/767}{Hydrodynamical simulations (767)}; \href{http://astrothesaurus.org/uat/600}{Galaxy interactions (600)}; \href{http://astrothesaurus.org/uat/416}{Dwarf galaxies (416)}}

\section{Introduction} \label{sec:intro}

The LMC is the most massive satellite galaxy of the Milky Way (MW), and has a barred spiral morphology. The LMC's stellar bar has several unusual properties that are not typical of galactic bars in the local universe. For example, the bar is offset from the center of the outer disk (R $\approx$ 5 kpc) by almost a kpc \citep{deVFreeman72, vdM2001, Zasov2002, Rathore2025} and resides in a different plane \citep{Haschke2012}, which is tilted with respect to the disk plane by $5^\circ - 15^\circ$ \citep{Choi2018a, Arranz2025}. The bar is neither evident in the spatial distribution nor the velocity fields of gas \citep{Stavely-Smith2003, Olsen2007}. Typically, in gas rich galaxies, bars drive gas inflows through angular momentum transport \citep[e.g.][]{Athanassoula1992, Kim2012, Beane2023, Romeo2023, Liang2024}, which is observable as streaming motions in the spatial distribution and velocity fields of both neutral and ionized gas \citep[e.g.][]{Bosma1978, Fathi2005, Erroz-Ferrer2015, Lopez-Coba2022}. This inflow is evident even in the case of offset, co-planar bars \citep{Athanassoula1989}.  

Early works suggested that the LMC bar might actually be an un-virialized structure above the plane of the disk \citep{ZhaoEvans2000}, or a manifestation of viewing a triaxial stellar bulge \citep{Zaritsky2004}. Recently, \citet[hereafter R25]{Rathore2025} accurately measured the LMC bar's parameters like radius (semi-major axis) and dynamical strength with completeness-corrected Gaia DR3 observations of red clump stars. They found that the LMC bar's properties are consistent with traditional barred galaxies in the local universe from the viewpoint of scaling relations between the bar radius, bar strength and the galaxy luminosity \citep{Erwin2005, Sheth2008, Guo2019, Cuomo2020}. Given that the LMC bar's strength and radius are similar to traditional barred galaxies, the origin of its unusual properties (offset, tilt and lack of gas inflows) is a pressing question. 

Numerical simulations suggest that the LMC bar's offset is likely a consequence of its interactions with its most prominent satellite, the SMC \citep[e.g.][]{Besla2012, Pardy2016, KRATOS2024}. However, the origin of the bar's tilt and the bar's inability to drive gas inflows is still not clear. Over the last decade, there have been several observations that support a recent ($\sim$ 100 Myr ago) collision (impact parameter $\sim$ 2 kpc) between the LMC and SMC \citep{Besla2016, Zivick2018, Zivick2019, Choi2022, Dhanush2024}. In this work, we perform a detailed characterization of the LMC's bar in an LMC-SMC collision scenario with the goal of testing the hypothesis that it is this recent collision that drives the LMC bar's offset and tilt. Further, we provide a framework to quantitatively compare the LMC's central gas distribution in different LMC-SMC interaction scenarios, which can be generalized to more advanced hydrodynamic simulations.

In addition to the bar tilt and offset, the LMC bar's pattern speed has also been a subject of constant debate. Using the spatial distribution of LMC's young star clusters, \cite{Dottori1996} inferred a bar pattern speed of $\Omega_b = 13.7 \pm 2$ km s$^{-1}$ kpc$^{-1}$. By modeling the asymmetric distribution of gas and star-formation in the LMC, \cite{Gardiner1998} inferred $\Omega_b \approx 40$ km s$^{-1}$ kpc$^{-1}$. \cite{Shimzu2012} assumed that the LMC's Shapley Constellation III star-forming region \citep{Shapley1951} is located at the L4 Lagrange point of a non-axisymmetric bar potential, and inferred $\Omega_b = 21 \pm 3$ km s$^{-1}$ kpc$^{-1}$. More recently, \cite{Arranz2024} measured the pattern speed of the LMC's bar using Gaia DR3 kinematics \citep{Luri2021}. Using the method of \cite{Dehnen2023}, they find a stationary or a slightly counter-rotating bar ($\Omega_b = -1.0 \pm 0.5$ km s$^{-1}$ kpc$^{-1}$). \cite{Kacharov2024} applied Schwarzschild Orbit Modeling \citep{Schwarzschild1979} to the LMC's inner regions and infer $\Omega_b = 11 \pm 4$ km s$^{-1}$ kpc$^{-1}$. 

Accurate determination of the LMC bar's pattern speed is important to place the LMC in context with other barred galaxies of the local universe and for studying bar-driven secular evolution in the LMC. However, the pattern speed of the LMC's bar is likely significantly affected by the recent SMC collision, which makes the observations challenging to interpret. In this work, we aim to build a framework to interpret the LMC bar's pattern speed in light of a recent SMC collision. 

Interactions between a larger disk galaxy and a smaller companion ($\sim 1:10$ mass ratio) is a common phenomenon in the local universe \citep{Zaritsky1993, Zaritsky1997, Besla2018, Chamberlain2024}, and has a significant effect on the evolution of barred galaxies. Such interactions can induce bar offsets \citep{Athanassoula1996, Athanassoula1997, Berentzen2003, Besla2012, Pardy2016, KRATOS2024}, affect the pattern speed of the bar \citep{Gerin1990, Sundin1991, Sundin1993, Athanassoula1997, Berentzen2003} and the distribution of gas in the bar region \citep{Noguchi1988, Berentzen2003, Besla2012}. When the companion is sufficiently massive and bound, interactions can also significantly weaken or destroy the bar \citep{Berentzen2003, Athanassoula2003a, KRATOS2024}. The final state of a bar in such interactions is sensitive to the mass of the companion \citep{Athanassoula1996, Athanassoula1997, Berentzen2003}. Excitingly, this implies that, if we establish that the LMC bar's unusual properties are primarily driven by the SMC, then we have an opportunity to constrain the SMC's total mass profile using the LMC's bar properties. 

We utilize hydrodynamic simulations of the LMC-SMC-Milky Way (MW) interaction history by \cite{Besla2012} for our investigation. \cite{Besla2012} present two scenarios, one where the LMC and SMC remain far away (with their closest separation being $>$ 20 kpc) which they refer to as Model 1; and the other where the LMC and SMC undergo a recent ($\approx$ 100 Myr ago) collision (impact parameter $\approx$ 2 kpc), which they refer to as Model 2. We analyze and compare the simulated LMC bar's properties in both models to assess whether a SMC collision is needed to explain the unusual bar.

We structure our manuscript as follows. In section \ref{sec:sims}, we describe the \cite{Besla2012} hydrodynamic simulation setup in some detail and further justify why it is appropriate to use for our study. In section \ref{sec:offset}, we analyze the simulated LMC bar's offset, and the role of the SMC's collision in driving the LMC bar's offset. In section \ref{sec:tilt}, we analyze the simulated LMC bar's tilt and the role of the SMC's collision in driving the bar's tilt. In section \ref{sec:ps}, we analyze the effect of the SMC's collision on the bar's pattern speed. In section \ref{sec:discussion}, we provide a framework to understand the influence of the SMC on the gas present in the LMC bar region. In the same section, we present a semi-analytic model to constrain the SMC's dark matter content by modeling its torques on the bar. We conclude in section \ref{sec:conclusion}. Throughout this manuscript, bold-italic mathematical font denotes vector quantities. Unit vectors are denoted by a hat symbol ( $\mathbf{\hat{}}$ ) over the vector. Quite often throughout the manuscript, we shall use the word \enquote{Clouds} to describe the LMC and SMC collectively.

\section{Description of the Hydrodynamic Simulations of the LMC-SMC-MW Interactions} \label{sec:sims}

In this section, we give a brief description of the N-body $+$ Smooth Particle Hydrodynamic (SPH) simulations of the LMC-SMC-MW interaction history used in this work; see \citet[hereafter B12]{Besla2012} for further details.

B12 modeled the interaction history of the Clouds over the past $6-7$ Gyr, including a MW infall for the past $1$ Gyr. The initial live dark matter halos of the LMC and SMC are modeled with Hernquist profiles, having a mass of $1.8 \times 10^{11}$ M$_\odot$ and $2.1 \times 10^{10}$ M$_\odot$, respectively, with a mass resolution per particle of $\approx10^6$ M$_\odot$. The initial live stellar disks of the LMC and SMC are modeled with exponential profiles having a total mass of $2.5 \times 10^{9}$ M$_\odot$ and $2.6 \times 10^{8}$ M$_\odot$, with a resolution per particle of $2500$ M$_\odot$ and $2600$ M$_\odot$, respectively. The initial SPH gas disks are modeled with exponential profiles with a total mass of $1.1 \times 10^{9}$ M$_\odot$ and $7.9 \times 10^{8}$ M$_\odot$, with a resolution per SPH cell of $3667$ M$_\odot$ and $2663$ M$_\odot$, respectively. The scale radii of the exponential disks for the LMC and SMC are 1.7 kpc and 1.1 kpc, respectively. The softening length for the star, gas and dark matter particles are 0.07 kpc, 0.07 kpc and 0.2 kpc respectively.

The B12 simulations have been performed with the Gadget-3 N-body SPH code \citep{Springel2005}. A subgrid multiphase model with an effective equation of state is used for the gas component, with radiative cooling following the \cite{Springel2003} prescription. The simulation includes star formation from cold gas using a Schmidt volume density law \citep{Springel2003}. Several works have used similar prescriptions to understand the evolution of cold gas in close galactic interactions \citep[e.g.][]{Berentzen2003, Hayward2014, Pardy2016, Pardy2018}. The simulated LMC's star formation history reasonably agrees with the observed constraints over the past 6 Gyr (B12).

As shown by \cite{Besla2010}, it is necessary to model the SMC as a disk to create the SMC's trailing stream of the required on-sky extent \citep[$\approx150^\circ$][]{Nidever2010}. Other studies like \cite{Pardy2018}, \cite{Luccini2024} have also adopted a similar approach for modeling the SMC. More details about the initial alignment of the SMC’s disk with the LMC-SMC orbit in the B12 simulations can be found in \cite{Besla2010, Besla2012}.

B12 have ensured that the initial disks and halos of the LMC and SMC are stable. They have verified that the density and velocity dispersion profile of the halos, density profile of the disks and the rotation curve of the disks do not significantly change in isolation. The LMC disk setup is chosen to be bar-unstable, and the bar forms out of secular evolution.

The LMC and SMC are allowed to interact with each other on an eccentric and decaying orbit. They were introduced in a static MW-like NFW potential \citep{NFW1997} corresponding to a virial mass of $1.5 \times 10^{12}$ M$_\odot$ and a concentration of 12 over the past 1 Gyr. The \enquote{present day} in the simulations was determined by matching the LMC's Galactocentric position and velocity to the observed values. 

B12 present two models - Model 1 and Model 2. In Model 1, the Clouds remain relatively far from each other, with their closest separation being $>25$ kpc. In Model 2, the SMC undergoes a direct collision with the LMC's disk (impact parameter $\sim 2$ kpc) around $100$ Myr ago. Here, the last pericenter of the SMC about the LMC occurred in the LMC's inner disk. Model 2 reproduces the observed structure and kinematics of the LMC significantly better compared to Model 1 \citep{Besla2016, Zivick2019, Choi2022, Rathore2025}, favoring a direct collision scenario. Thus, we use Model 2 as our primary simulation and Model 1 as a reference for comparison. Comparing Model 1 and Model 2 simulation will determine if a recent SMC collision is necessary for explaining the morphological peculiarities of the LMC's bar as opposed to effects of secular disk evolution or weak SMC tides.

We refer the reader to the B12 paper for a detailed comparison of the simulated external features of the Clouds with observations. These external features include the SMC's trailing gas stream \citep{Mathewson1974, Braun2004, Nidever2010}, the LMC-SMC bridge \citep{Kerr1957, Putman2003,  Bruns2005} and the present day Galacto-centric position and velocity of the LMC \citep{Kallivayalil2006a, Kallivayalil2006b}.

The B12 simulations are in reasonable agreement with several internal features of the LMC. This includes the LMC's one-armed spiral \citep{deVFreeman72}, the LMC's off-centered bar \citep{deVFreeman72, vdM2001, Rathore2025}, and the LMC's outer disk morphology \citep{Mackey2016, Besla2016}. However, the B12 scenario is just one plausible LMC-SMC-MW interaction history and should not be expected to accurately reproduce all of the prominent observables associated with the Clouds. In section \ref{sec:limitations}, we discuss how the limitations of the B12 simulations affects our study.

The LMC bar's properties at present day in B12 Model 2 were rigorously compared to observations by R25, and were found to be remarkably consistent. In particular, the simulated bar's dynamical strength, radius and the radial profile of the $m = 2$ Fourier component agree with observations within the statistical uncertainties. Thus, it is reasonable to study the physics of the simulated bar in the B12 simulations as a representation of the real system.

\subsection{The Geometry of the SMC's Orbit Around the LMC} \label{sec:lmc_smc_geo}

\begin{figure}
\centering
	\includegraphics[width=\columnwidth]{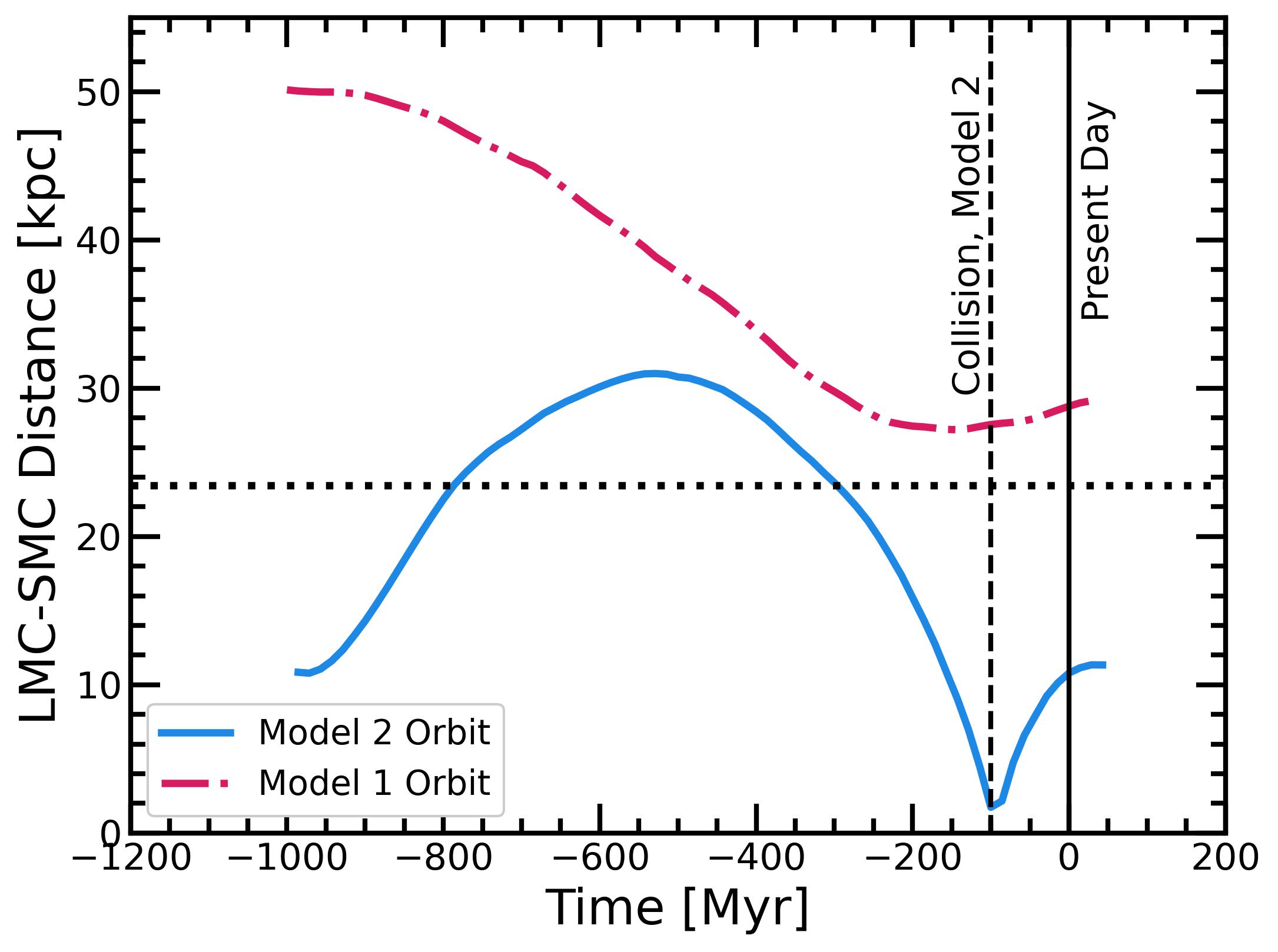}
    \caption{The distance between the stellar center of mass of the LMC and SMC for the B12 Model 1 (magenta dash-dot line) and Model 2 (blue solid line) simulation, as a function of time. The solid black line denotes the \enquote{present day} in the simulation, which is $\approx$~1 Gyr after the Clouds cross the virial radius of the MW's halo. In Model 2, the SMC collides with the LMC (impact parameter $\approx$ 2 kpc) at the epoch denoted by the black dashed line. The collision occurred $\approx$ 100 Myr before the present day. In Model 1, the Clouds remain far from each other with their closest separation being $> 25$~kpc. The horizontal dotted line shows the observed separation between the Clouds. In section \ref{sec:limitations}, we discuss how the difference between the simulated and observed separation at present day affects our conclusions.}
    \label{fig:orbit}
\end{figure}

\begin{figure*}
    \centering
    \includegraphics[width=\linewidth]{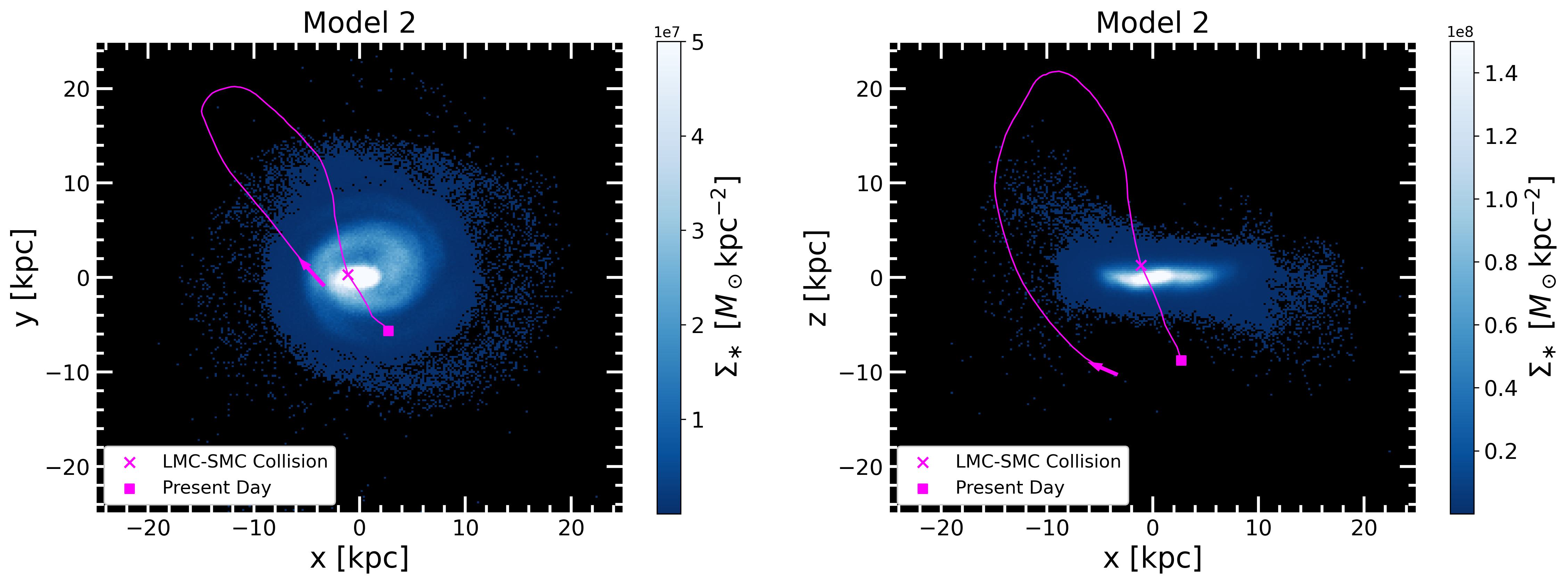}
    \caption{The visualization of the SMC's orbit about the LMC for the Model 2 simulation, plotted in the LMC's frame of reference over the past 1 Gyr as the galaxies orbit the MW. The left (right) panel shows the face-on (edge-on) projection of the surface density ($\Sigma_\ast$) of the LMC's simulated stellar disk at the present day. The magenta curve shows the orbit of the SMC's stellar center of mass with the arrow indicating the position and direction of SMC's motion 1 Gyr ago. The magenta cross marks the location of the SMC when the two galaxies collide ($\approx$100 Myr ago, impact parameter of $\approx 2$ kpc). The magenta square marks the location of the SMC at present day in the simulation. On its closest approach to the LMC, the SMC's orbital plane is inclined to the LMC's disk plane by around 50$^\circ$. With such an orbital geometry, the SMC can affect both the vertical and the in-plane motion of the LMC's bar.}
    \label{fig:orbit_vis}
\end{figure*}

\begin{figure*}
    \centering
    \includegraphics[width=\linewidth]{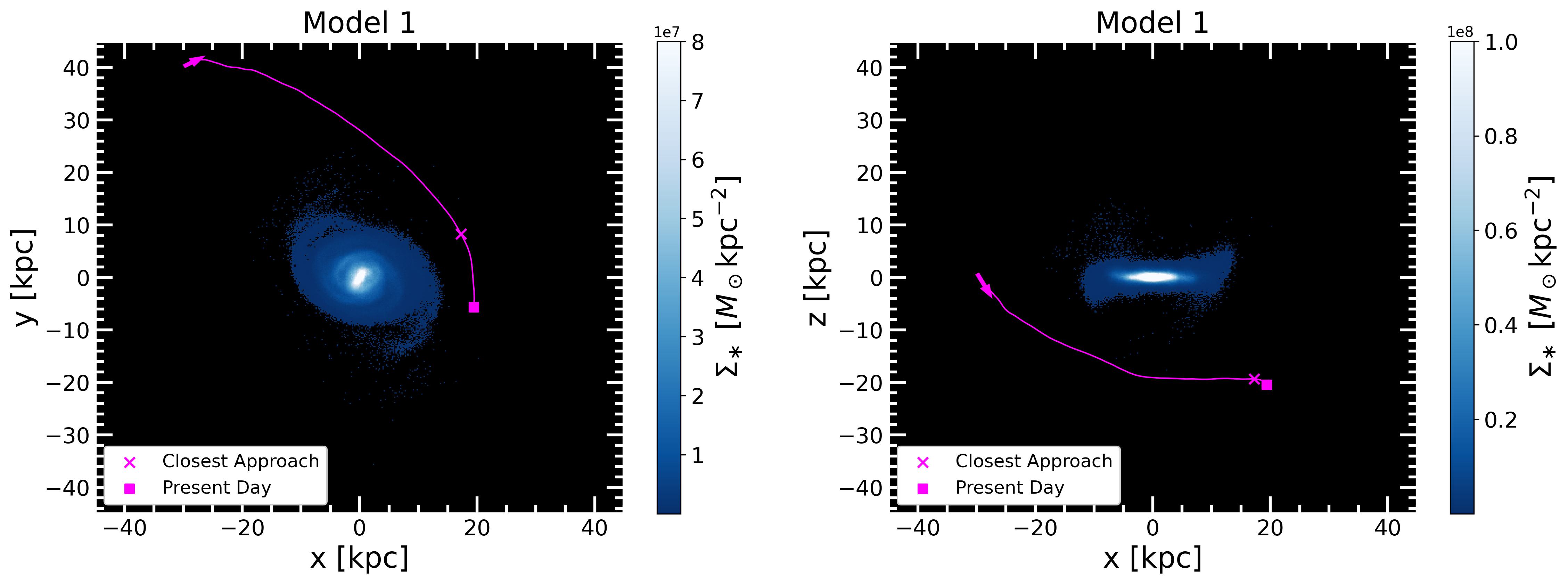}
    \caption{Same as Figure \ref{fig:orbit_vis}, but for the Model 1 simulation. In Model 1, the LMC and SMC remain much farther away as compared to Model 2 (note the difference in axis scales from Figure \ref{fig:orbit_vis}). The magenta cross marks the location of the SMC at its closest approach to the LMC ($\approx$ 27 kpc). In Model 1, we do not expect the SMC to significantly perturb the LMC's bar due to the much larger separation between the two galaxies as compared to Model 2 (where the closest approach is $\approx$ 2 kpc). We use Model 1 as a control simulation for comparison with the bar properties of Model 2.}
    \label{fig:orbit_vis_m1}
\end{figure*}

The distance between the stellar center of mass of the LMC and SMC over the MW infall duration for Model 1 and Model 2 simulations is shown in Figure \ref{fig:orbit}. In Model 2, the closest separation ($\approx$ 2 kpc) between the LMC and SMC was approximately 100 Myr ago, which was also a direct collision between the two galaxies (the SMC's pericenter resided in the LMC's disk). On the other hand, the closest separation between the Clouds in Model 1 is ${\approx}~27$~kpc. We outline our method for computing the center of mass of the LMC and SMC for each simulation snapshot in section \ref{sec:com}. In Figure \ref{fig:orbit}, we also show the observed separation between the Clouds \citep[$\approx 20$ kpc,][]{Kallivayalil2006a, Kallivayalil2006b}. The simulated separation at present day is smaller than the observed separation by $\approx 10$ kpc. In section \ref{sec:limitations}, we shall discuss how this difference affects our conclusions.

Next, we show the geometry of the encounter between the LMC and SMC in the Model 1 and Model 2 simulations. In particular, we compute the orientation of the orbital plane of the SMC relative to the LMC's disk plane, since that allows us to understand in what ways the SMC can affect the LMC's bar. For instance, if the SMC's orbital plane is closely aligned with the LMC's disk plane, the SMC can primarily affect the bar's in-plane motions. With such an orientation, the SMC can potentially induce bar offsets as well as slow down/speed up the bar, but will not be able to significantly affect the bar's vertical degrees of freedom, like tilt. On the other hand, if the SMC's orbital plane is significantly misaligned with respect to the LMC's disk plane, then the SMC can affect the bar's in-plane and vertical degrees of freedom.

We quantify the orientation of the SMC's orbit relative to the LMC's disk plane by computing the direction of the SMC's orbital angular momentum about the LMC's stellar center as a function of time. Let $\vect{v_{SMC}}$ and $\vect{R_{SMC}}$ be the velocity and position vectors of the SMC's center relative to the LMC's center. We compute the direction of the specific orbital angular momentum of the SMC (denoted by $\vect{\hat{L}_{SMC}}$) as follows:

\begin{equation}
    \vect{\hat{L}_{SMC}} = \vect{\hat{R}_{SMC}} \times \vect{\hat{v}_{SMC}}
\end{equation}

Then, we compute the angle between the orbital plane of the SMC and the LMC's disk as follows:
\begin{equation}
    i = \arccos{(\vect{\hat{L}_{SMC}} \cdot \vect{\hat{J}_{LMC}}}),
\end{equation}
\noindent where $\vect{\hat{J}_{LMC}}$ is the direction of angular momentum of the LMC's disk, which is computed by averaging the angular momentum vector of each disk star within 10 kpc of the LMC's center in each snapshot. We have rotated the coordinate frame such that $\vect{\hat{J}_{LMC}}$ is aligned with the Z-axis.

A completely polar orbit where the SMC is orbiting perpendicular to the LMC's disk plane would result in $i = 90^\circ$. On the other hand, a completely equatorial orbit, where the SMC is orbiting in the same plane as the LMC's disk, would result in $i = 0^\circ$. We find that the orientation of the SMC's orbital plane relative to the LMC's disk has a significant time dependence in Model 2. At the epoch where the LMC-SMC system first crosses the virial radius of the MW (around 1 Gyr ago), the SMC's orbit is predominantly polar relative to the LMC's disk plane ($i = 74^\circ$). The SMC's orbit becomes more equatorial with time as it approaches the LMC within the tidal field of the MW. When the two galaxies collide, $i = 52^\circ$, indicating that the SMC's orbit has significant equatorial and polar components. Hence, the SMC can affect both in-plane and vertical motions of the LMC's bar.

Figures \ref{fig:orbit_vis} and \ref{fig:orbit_vis_m1} show the SMC's orbit about the LMC in the LMC's frame of reference for Model 2 and Model 1 respectively. We show both the face-on and edge-on projections of the LMC's disk.

\subsection{Obtaining the Stellar Center of Mass of the LMC and SMC} \label{sec:com}

Computing the correct stellar center of mass is important to get an accurate distance between the simulated LMC and SMC as a function of time. We use the method outlined in \cite{Power2003}, which is based on an iterative shrinking sphere algorithm. First, we obtain an approximate center of mass of the LMC and SMC by calculating the mass weighted average position of the stars constituting the LMC and SMC respectively. Then, we define a sphere of radius 10 kpc (5 kpc) centered on this mass-weighted average position of the LMC (SMC), and compute the center of mass of stars that reside inside the spheres. Then, we shrink the radius of the spheres by $30\%$, and recompute the center of mass of the stars that reside within the shrunken spheres. We repeat until a convergence criterion has been achieved, wherein the center of mass does not change by more than 0.01 kpc between two successive shrinking iterations. This method has been used before for both numerical simulations \citep{Power2003, GC2019} and resolved stellar populations in observations (R25). A principal advantage of this method is that the inferred center of mass usually converges to the center of mass found by weighting the position of each star by the gravitational potential at that location \citep{Power2003}. Hence, the inferred center of mass is not biased by tidal perturbations in the outskirts.

\section{The LMC Bar's Offset} \label{sec:offset} 

\begin{figure*}
    \centering
    \includegraphics[height = 0.3\textwidth, width=0.3\textwidth]{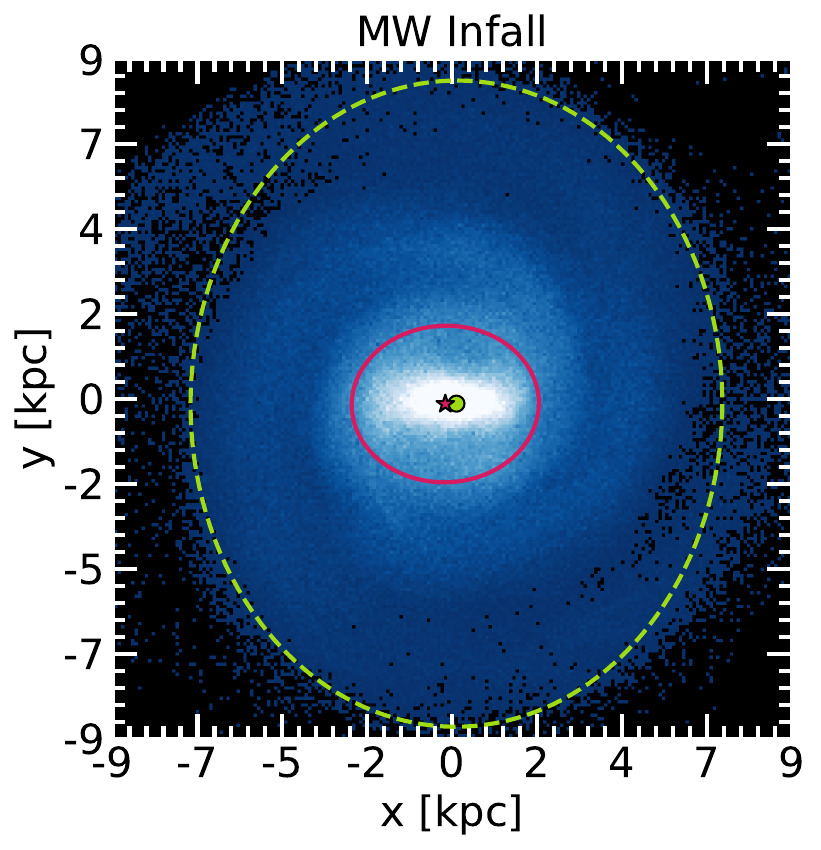}
    \includegraphics[height = 0.3\textwidth, width=0.3\textwidth]{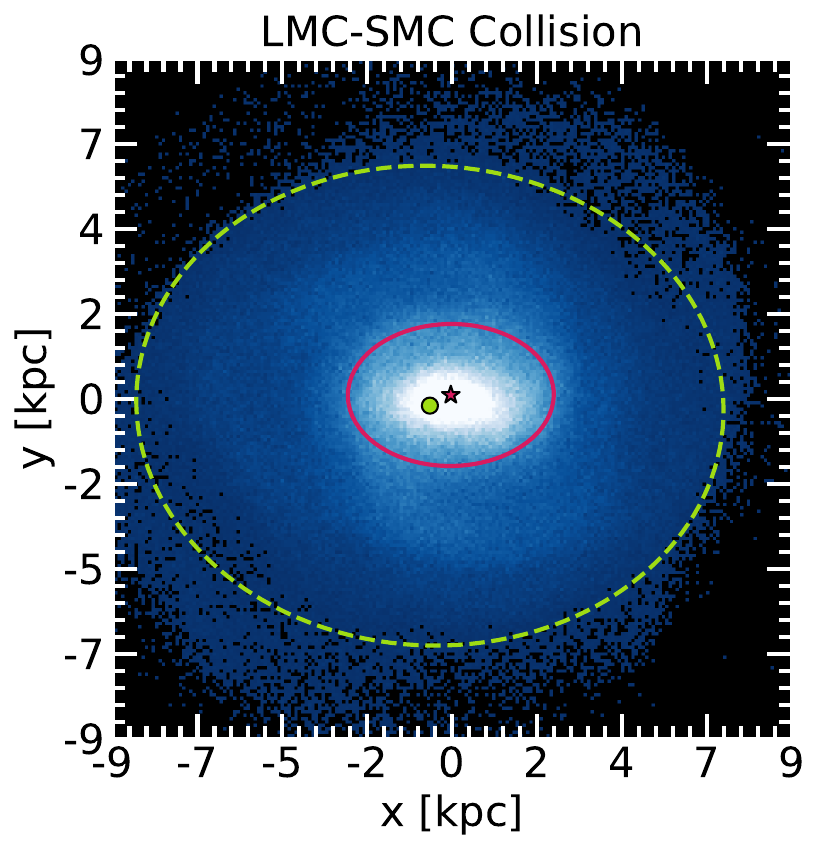}
    \includegraphics[height = 0.3\textwidth, width=0.35\textwidth]{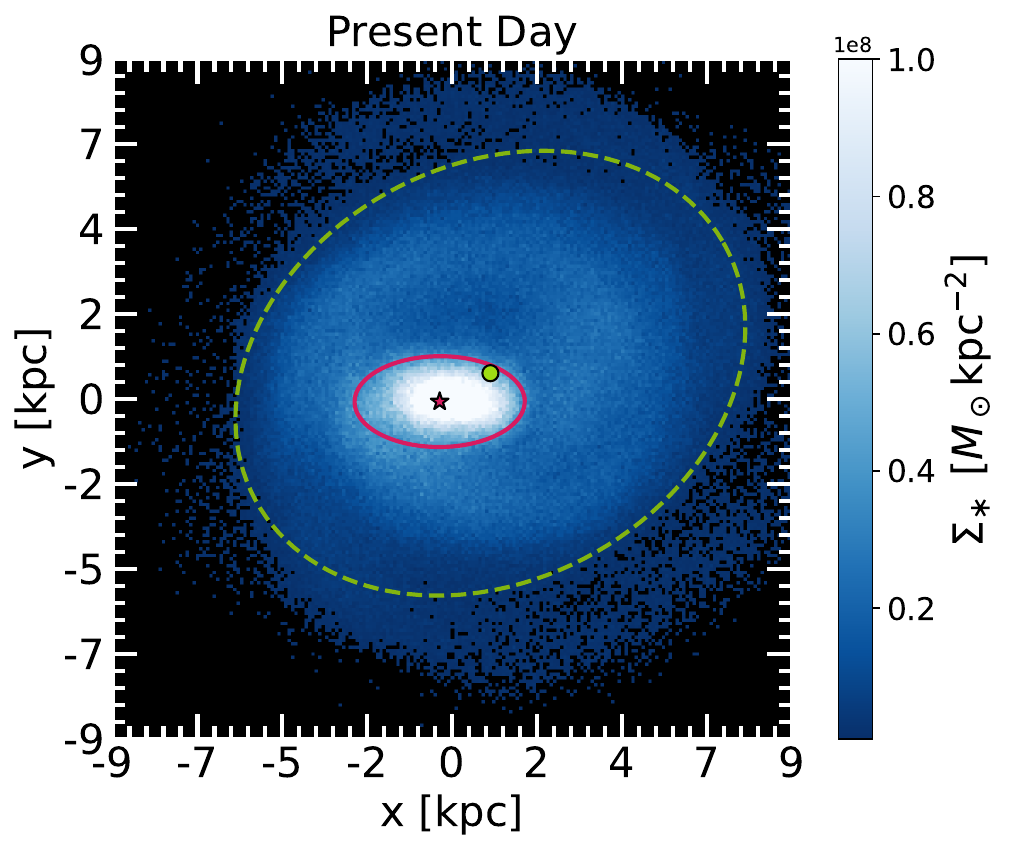}
    \caption{The offset of the simulated LMC bar at the epoch of infall into the MW (left panel), LMC-SMC collision (middle panel) and present day (right panel) is illustrated using the surface density of stars ($\Sigma_\ast$) in the LMC disk. The LMC stellar disk is plotted in the XY plane in the LMC's frame of reference. The solid-magenta ellipse (bar ellipse) is an isodensity elliptical contour with a semi-major axis equal to the bar radius. The magenta star is the geometric center of the bar ellipse. The dashed-green ellipse (outer disk ellipse) is an isodensity elliptical contour with a semi-major axis equal to the radius where the surface density profile of the disk drops by a factor of 100. The green circle is the geometric center of the outer disk ellipse. The separation between the center of the bar ellipse and the center of the outer disk ellipse is defined as the bar offset. The LMC's bar develops a large offset ($\approx 1.5$ kpc) at present day. The offset is small at the epochs of MW infall and the LMC-SMC collision, indicating that the LMC's bar develops a large offset post SMC collision. The small offset at the MW infall epoch is equivalent to the offset found in Model 1 at the present day (where the Clouds do not collide).}
    \label{fig:ellip_offset}
\end{figure*}

\begin{figure}
    \centering
    \includegraphics[width=\columnwidth]{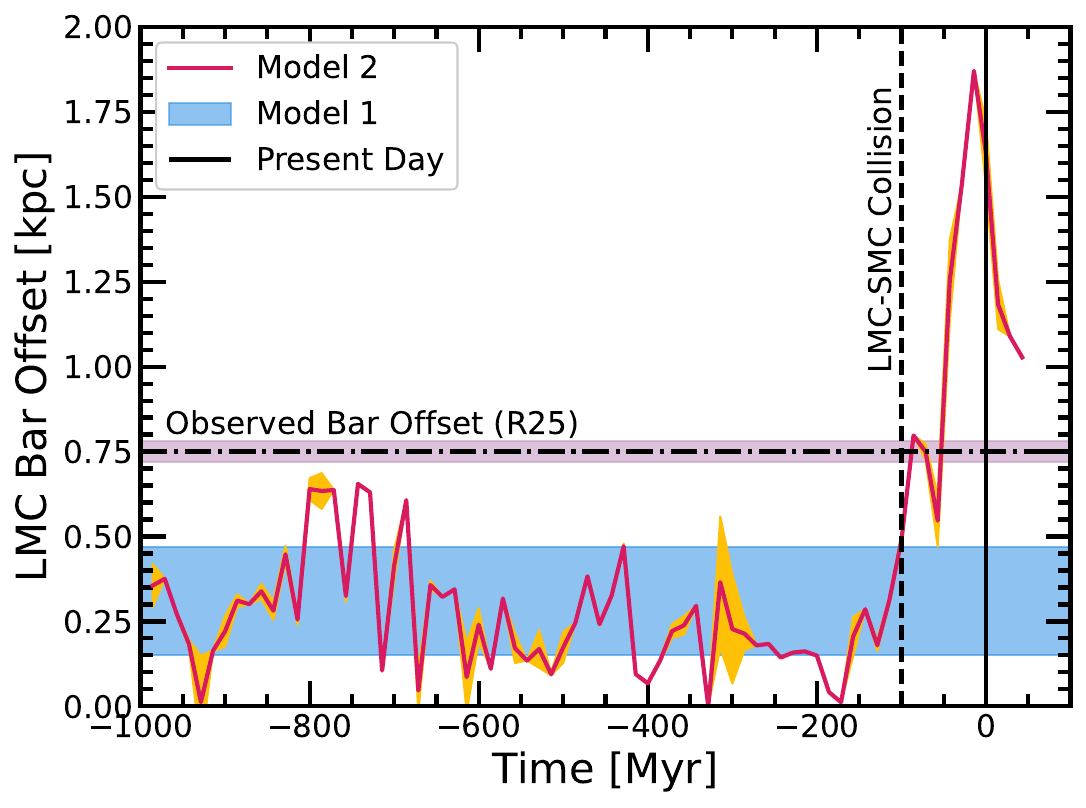}
    \caption{The LMC bar's offset as a function of time. The red solid line is the measured bar offset in the Model 2 simulation and the shaded yellow band around it is the measurement error. The black dash-dot line marks the observed bar offset of 0.76 kpc (R25) with the purple shaded band being the $3-\sigma$ error on the observations. The vertical black dashed line marks the LMC-SMC collision epoch and the vertical black solid line marks the present day epoch in Model 2. The LMC's bar develops a large offset just as the SMC collides with the LMC. The present-day offset is larger than observed, suggesting that the true collision must have occurred 150-200 Myr ago such that the bar offset has sufficiently decayed, but is still present. The blue shaded band denotes the $1-\sigma$ spread on either side of the mean bar offset measured in the Model 1 simulation. In Model 1, where the LMC and SMC remain far away, the bar does not develop an offset to an extent seen in Model 2 and observations at the level of $7-\sigma$ and $3-\sigma$ respectively in statistical significance.}
    \label{fig:offset_time}
\end{figure}

We measure the bar offset from the outer disk center as a function of time in Model 2 using the framework of R25. First, we center each snapshot on the LMC's stellar center of mass as computed in \ref{sec:com}, and rotate the coordinate system such that the LMC's disk is aligned with the XY plane (see section \ref{sec:lmc_smc_geo}). Then, we compute the bar radius (semi-major axis) $R_{bar}$ and the bar position angle $\Phi_{bar}$ for each snapshot using Fourier decomposition \citep{AM2002, Lucey2023, Silva2023, Ghosh2024}.

We convert the position of each star from Cartesian coordinates (x, y) to polar coordinates (R, $\omega$). We bin the disk in radial annuli with a bin size of 0.1 kpc. In each bin, we perform a Fourier transform in the azimuthal coordinate to obtain a cosine series and a sine series: 

\begin{equation} \label{eq:alpha}
\alpha_m (R) = \frac{\sum_i M_i \cos(m \omega_i)}{N}, \qquad m = 0, 1, 2, ... 
\end{equation}

\begin{equation} \label{eq:beta}
\beta_m (R) = \frac{\sum_i M_i \sin(m \omega_i)}{N}, \qquad m = 1, 2, 3, ...
\end{equation}
\noindent where $M_i$ is the mass of the star-particle, the integer $m$ is the spatial frequency harmonic and the index $i$ runs over all the stars in a given radial bin.

The Fourier phase is given by:
\begin{equation} \label{eq:phi}
    \Phi_m (R) = \frac{1}{m}\arctan\left(\frac{\beta_m}{\alpha_m}\right), \qquad m = 1, 2, 3, ...
\end{equation}

The bar is a rigid bi-symmetric structure that is described by the $m = 2$ component of the above Fourier series. $\Phi_2$ describes the phase of the bar, and is expected to be constant in the bar region and then deviate from a constant near the end of the bar. Following R25, we define the end of the bar as the radius $R_{bar}$ where $\Phi_2(R)$ deviates from a constant by $10^\circ$. For the present day snapshot in Model 2, $R_{bar} = 2.20$ kpc. The error in the bar radius is dominated by the softening length of the simulation, which is 0.1 kpc. We will propagate this error to the bar offset.

We compute $\Phi_{bar}$ from the $m = 2$ phase angle of all stars that reside within 3 kpc of the center:
\begin{equation} \label{eq:pos_angle}
    \Phi_{bar} = \Phi_2 (R < 3 \: \: \rm{kpc})
\end{equation}
Using $\Phi_{bar}$, we align the bar with the X-axis in all the snapshots. Hereafter, the coordinate frame centered at the LMC's center of mass where the LMC's disk is aligned with the XY plane and its bar is aligned with the X-axis will be referred to as the \enquote{LMC bar frame}. 

Following R25, we fit iso-density ellipses to the disk surface mass density distribution in each snapshot with the python package {\em photutils.isophote.Ellipse}, which uses the algorithm of \cite{Jed1987}. We define the bar ellipse as the iso-density ellipse whose semi-major axis equals $R_{bar}$. Thus, the bar ellipse traces the density at the end of the bar. We define the bar center as the geometric center of the bar ellipse. 

We define the outer disk ellipse as the iso-density ellipse whose semi-major axis equals the outer disk radius. Following R25, we calculate the outer disk radius as the radius where the surface density profile of the disk drops by a factor of 100 relative to its central value. To characterize the surface density profile of the disk, we fit an exponential model to the particle distribution using the python package {\em scipy.optimize.curvefit}:
\begin{equation} \label{eq:exp_profile}
\Sigma(R) = \Sigma_0 e^{\frac{-R}{R_s}}
\end{equation}
\noindent where $\Sigma(R)$ is the surface density of stars in a given radial bin. $\Sigma_0$ and $R_s$ are the fit parameters. $R_s$ is the exponential scale radius of the disk. The outer disk radius ($R_{od}$) is given by:
\begin{equation} \label{eq:R_od}
R_{od} = R_s \: \ln(100)
\end{equation}

For the present day in Model 2, the outer disk radius is 7.43 kpc. We define the outer disk center as the geometric center of the outer disk ellipse. The bar offset is given by the distance between the bar center and the outer disk center. 

We calculate the error in the bar offset by computing the separation between the centers of the iso-density ellipses whose semi-major axis is one softening length different from $R_{bar}$. 

Figure \ref{fig:ellip_offset} shows the bar ellipse and the outer disk ellipse along with their centers for three epochs in Model 2 - when the LMC and SMC are at the MW virial radius (MW infall epoch, left panel), when the LMC and SMC collide (LMC-SMC collision epoch, middle panel) and the present day epoch (right panel). We see a clear separation between the bar center and the outer disk center in the present day snapshot, indicating a significant bar offset. The separation between the bar center and the outer disk center is less at the LMC-SMC collision epoch and is negligible at the MW infall epoch. Thus, the LMC's bar develops a strong offset from the outer disk center post SMC collision, and the bar is mostly aligned with the outer disk center prior to the SMC collision. The offset at the MW infall epoch of Model 2 is representative of the typical offset seen in Model 1 (where the Clouds do not collide), indicating that a strong interaction between the Clouds is needed to generate a pronounced bar offset. 

In Figure \ref{fig:offset_time}, we plot the LMC bar's offset as a function of time in Model 2. We mark the epoch of LMC-SMC collision as a dashed black line, and the present day epoch as a solid black line. The bar offset increases significantly just as the SMC collides with the LMC, indicating a strong causal connection between these two events. 

The observed LMC bar offset ($0.76 \pm 0.01$ kpc, R25) is shown as a horizontal dashed line in Figure \ref{fig:offset_time}. The offset at present day in the simulation ($\approx$ 1.5 kpc) is significantly larger compared to observations. However, the simulated offset evolves substantially with time after the collision. In the future, it is expected that the offset will diminish and the disk and bar centers will be coincident. The simulated LMC needs to evolve for {\em at least} 50 Myr (i.e. till the end of the simulation) after the simulation present day for the bar offset to be consistent with observations. This suggests the LMC-SMC collision to have happened {\em at least} 150 Myr ago. 

This impact timing is consistent with \cite{Choi2022}, who compared the kinematics of the LMC's disk in Model 2 with Gaia EDR3 observations and concluded that the Model 2 disk needs to evolve for at least 50 Myr more after the simulation present day for the simulated kinematics to match observations. 

Another possible solution for matching the bar offset in Model 2 to observations is if the simulated present day was around 50 Myr earlier, which would place the LMC-SMC collision 50 Myr ago. However, the Clouds would be very close to each other today (a few kpc separation) in such a scenario, which is inconsistent with observations. 

The simulation ends before the bar and disk centers are coincident. Assuming that the bar offset continues to decay at a similar rate as simulated, the bar will remain offset for a total of $\approx200$ Myr post-collision. Thus, if the collision happened more than 200 Myr ago, we should not observe any offset at present day. This suggests the LMC-SMC collision to have happened {\em at most} 200 Myr ago. Hence, based on the observed bar offset, we suggest the LMC-SMC collision to have happened 150-200 Myr ago.

We also compute the bar offset in Model 1 as a function of time. The offset in Model 1 is significantly smaller compared to Model 2 at all times. We include the mean offset (0.31 kpc) and the $1-\sigma$ spread (0.16 kpc) in Model 1 as a blue shaded band in Figure \ref{fig:offset_time}. Model 1 is not able to reproduce the observed bar offset and the offset in Model 2 at a level of $3-\sigma$ and $7-\sigma$ respectively in statistical significance.  

Using idealized N-body simulations, \citet[][hereafter APB97]{Athanassoula1997} studied the behavior of a disk galaxy that collides with a smaller spherical companion. They considered disk scale lengths ranging from 2.5 kpc to 15 kpc, mass ratios ranging from 0.02 to 0.2, impact parameters ranging from $\approx0$ kpc to 3.6 kpc, and impact angles ($i$) ranging from $41^\circ$ to $89^\circ$. They found that in almost all scenarios the bar gets offset from the outer disk center. Their offsets lasted 200-300 Myr, which is consistent with our findings. 

\citet[][hereafter B03]{Berentzen2003} investigated the consequences of a vertical impact ($i \approx 90^\circ$) of a smaller spherical companion (mass ratio 1:5) on a gas-rich barred disk through idealized hydrodynamic simulations. Their disk scale length was 3 kpc. They considered three scenarios: a central impact, an impact along the bar major axis (6 kpc from the disk center), and an impact along the bar minor axis (3 kpc from the disk center). They found their bar offsets lasted for $\approx 600$ Myr which is longer than ours. However, they had larger bar offsets (2.5 - 3 kpc) compared to ours (1 - 1.5 kpc) presumably due to their larger impact parameters and a larger mass ratio, so it is not surprising that their offsets took more time to reduce. 

\citet[hereafter JA24a]{KRATOS2024} built a library of full N-body interaction scenarios for LMC, SMC and MW like galaxies, and found their maximum bar offsets to be in the range of 2 - 3 kpc, lasting for $\approx 500$ Myr. However, their closest separation between the Clouds ($\approx$ 6 kpc) was larger than in B12 Model 2 ($\approx$~2~kpc). 

We conclude that our finding of a bar offset in the LMC that is induced by a direct SMC collision is consistent with studies in the literature.  

\subsection{Separation between the LMC's Stellar and Dark Matter centers of mass}

\begin{figure*}
    \centering
    \includegraphics[width=0.49\textwidth]{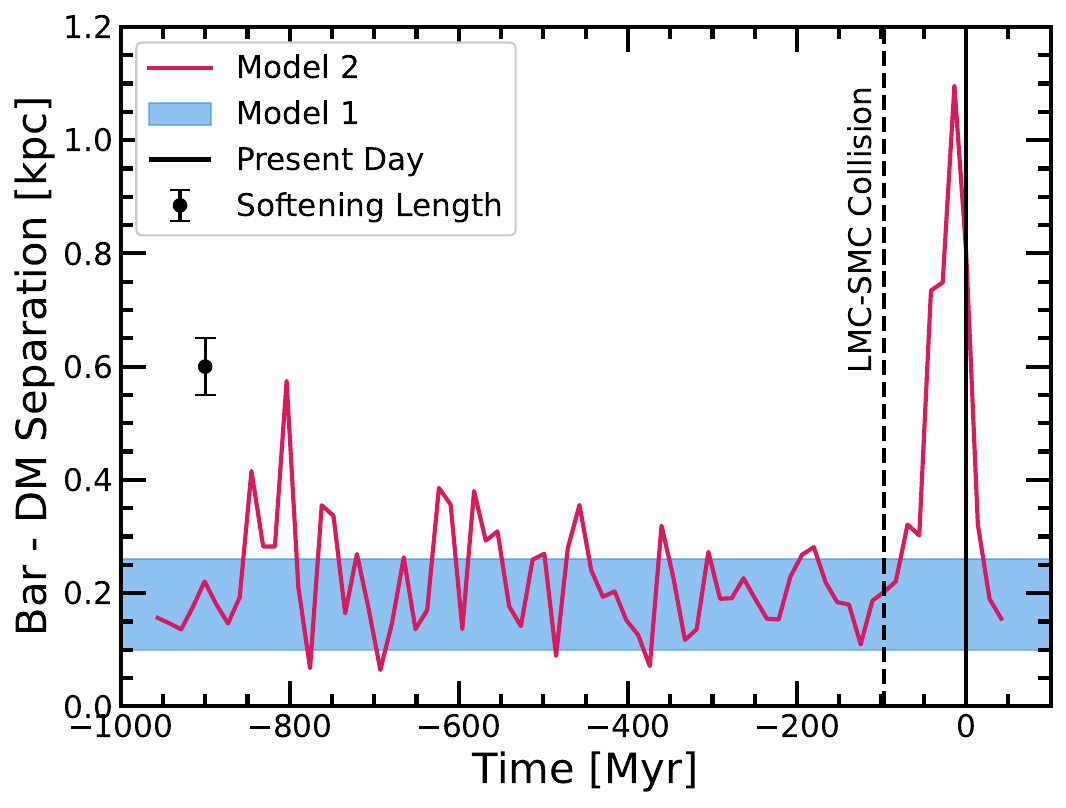}
    \includegraphics[width=0.49\textwidth]{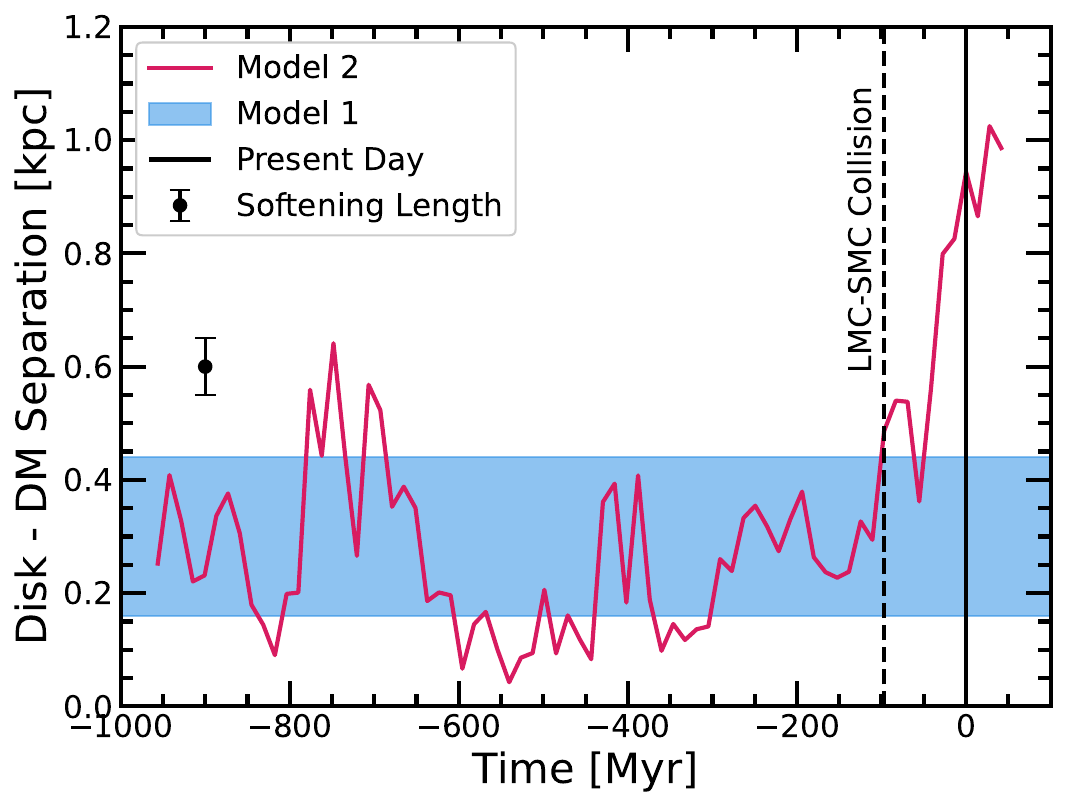}
    \caption{The separation between the simulated LMC's bar center (left) and  outer disk center (right) from the dark matter (DM) center of mass (COM) as a function of time. {\em Left panel:} The red solid line denotes the bar - DM separation for the Model 2 simulation. The separation increases just as the SMC collides with the LMC (black dashed line), becoming $\approx1$ kpc at the simulated present day (black solid line). This is coincident in time with the bar becoming offset from the outer disk center (Figure \ref{fig:offset_time}). However, the bar - DM separation decays faster than the bar offset, becoming negligible 150-200 Myr after the collision. The blue shaded band denotes the $1-\sigma$ spread on either side of the mean bar - DM separation as measured in Model 1. In Model 1, where the Clouds do not collide, the separation is comparable to the simulation softening length (black error bar) and remains significantly smaller than Model 2 throughout the simulation. {\em Right Panel}: The red solid line denotes the outer disk - DM separation for the Model 2 simulation. Contrary to the bar - DM separation, a significant outer disk - DM separation ($\approx$ 1 kpc) is expected in the observed LMC today.}
    \label{fig:bar_dm_offset}
\end{figure*}

In this section we investigate the time evolution of the simulated LMC's dark matter (DM) center of mass (COM) relative to the geometric center of the bar and outer disk during its interaction with the SMC. This is important to understand the state of disequilibrium of the LMC galaxy. 

We compute the DM COM by applying the iterative shrinking process described in section \ref{sec:com} to the DM particles, with an initial radius of $\approx 120$ kpc (the chosen virial radius of the LMC) for the sphere \footnote{We have tried different values for the initial radius and the shrinking fraction for the shrinking sphere method and find differences of a few softening lengths in the DM COM inference. However, the main conclusions of this section do not change.}.

The left panel of Figure \ref{fig:bar_dm_offset} shows the separation between the bar center and the DM COM for Model 2. The bar center separates from the DM COM just as the SMC collides with the LMC. The maximum separation is $\approx 1$ kpc, around the simulation present day. However, the bar{-}DM separation decays significantly faster compared to the bar offset (Figure \ref{fig:offset_time}). Based on the timing of the LMC-SMC collision inferred from the bar offset (150-200 Myr ago, section \ref{sec:offset}), we do not expect any separation to exist between the bar center and the DM COM in the LMC today. The separation in Model 1, where the LMC and SMC do not collide, has a mean of 0.18 kpc with a standard deviation of 0.08 kpc. This is significantly smaller than the separation in Model 2, when the LMC and SMC collide.

The right panel of Figure \ref{fig:bar_dm_offset} shows the separation between the outer disk center (as defined in section \ref{sec:offset}) and the DM COM for Model 2. The outer disk center separates from the dark matter center by $\approx 1$ kpc just as the SMC collides with the LMC. This is coincident in time with the separation of the bar from the DM COM. However, the outer disk - DM separation does not decrease within the time span of the simulation. Hence, in the observed LMC, we  expect a significant separation ($\approx$ 1 kpc) to exist between the center of the outer-most disk isophote and the center of the LMC's DM halo, indicating that the LMC's outer disk is in a high state of dis-equilibrium with the DM halo.

Our finding is consistent with previous works like APB97 and B03 who find that the bar center separates from the DM COM as the bar gets offset from the outer disk center. \cite{Pardy2016} perform hydrodynamic simulations of the interaction between LMC and SMC like galaxies, and investigate both equatorial and polar SMC-LMC orbital configurations. They find that in all of their configurations, the bar center becomes separated from the DM COM. However, their separation lasts $\approx 2$ Gyr, which is significantly longer compared to ours. The bar offset seen in \cite{Pardy2016} simulations and the bar offset seen in the B12 simulations (and APB97, B03) is likely a result of a different dynamical process since they have such different timescales of decay, which requires further investigation. 

\section{The LMC Bar's Tilt} \label{sec:tilt}

\begin{figure*}
    \centering
    \includegraphics[height = 0.3\textwidth, width = 0.3\textwidth]{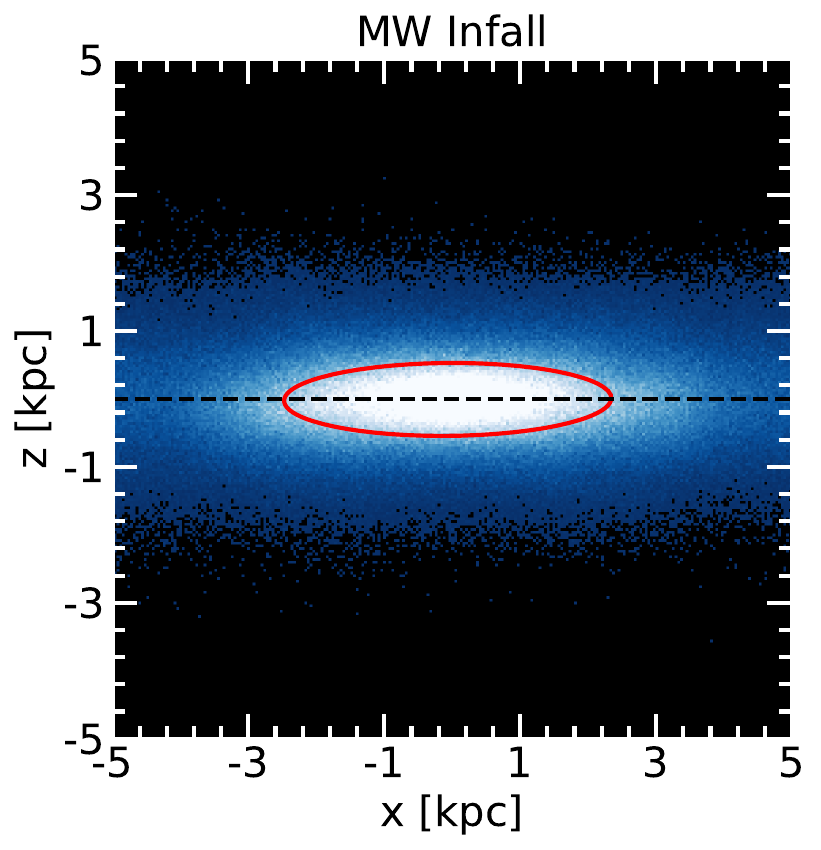}
    \includegraphics[height = 0.3\textwidth, width = 0.3\textwidth]{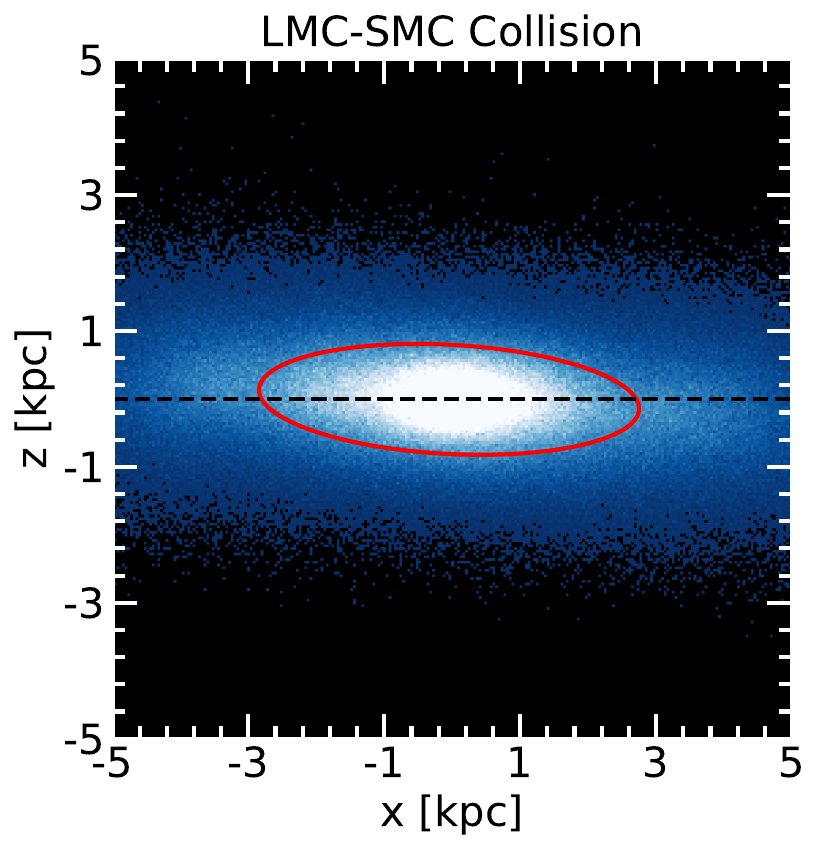}
    \includegraphics[height = 0.3\textwidth, width = 0.35\textwidth]{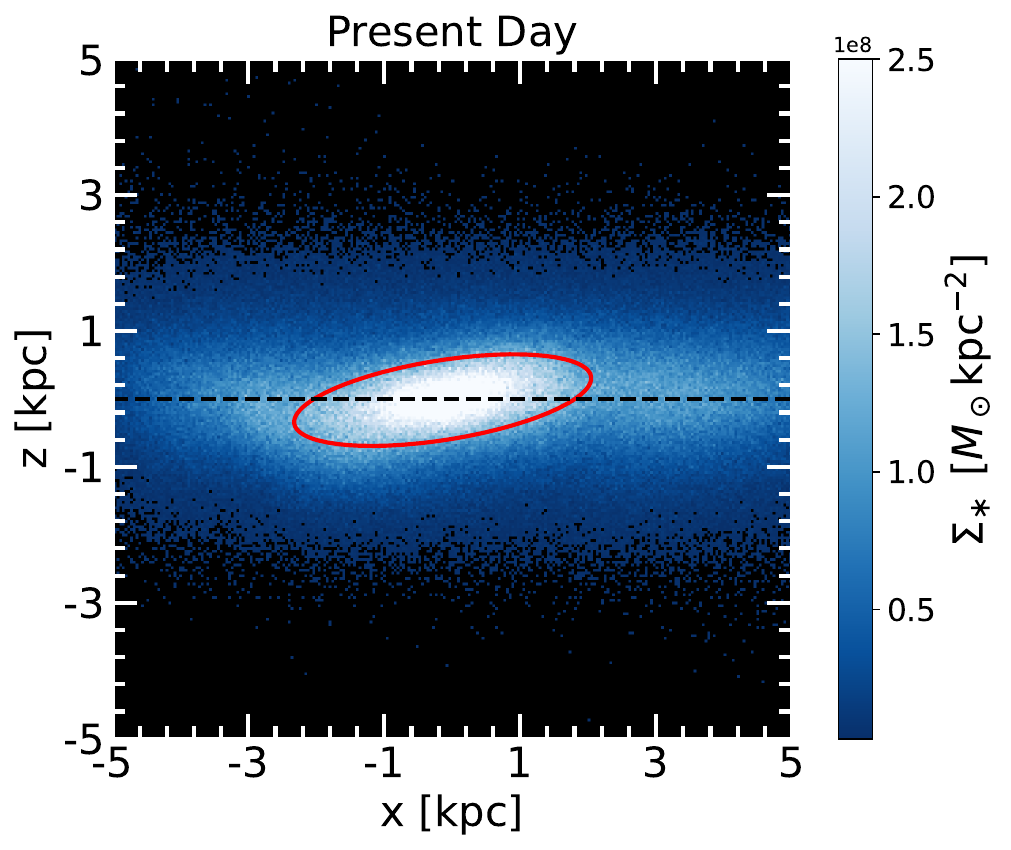}
    \caption{Visualization of the tilted bar and stellar disk of the Model 2 simulated LMC at the epoch of MW infall (left panel), LMC-SMC collision (middle panel) and present day (right panel). The LMC's stellar disk is plotted edge-on (XZ projection) in its center of mass frame. The colorbar shows the stellar surface density in the XZ projection ($\Sigma_\ast$). The red ellipse denotes the isodensity ellipse with a semi-major axis equal to the radius of the bar (section \ref{sec:offset}). The black-dashed line denotes the disk plane (Z = 0 in this frame of reference). As emphasized by the red ellipse, the simulated stellar bar is tilted with respect to the disk plane by $\approx$ 8.6 $^\circ$ at present day. The tilt is small at the epochs of MW infall and the LMC-SMC collision, indicating that the LMC's bar develops a strong tilt post SMC collision. The negligible tilt at the MW infall epoch is also representative of the tilt seen in the Model 1 simulation (where the Clouds do not collide).}
    \label{fig:ellipse_tilt}
\end{figure*}

\begin{figure}
    \centering
    \includegraphics[width=\columnwidth]{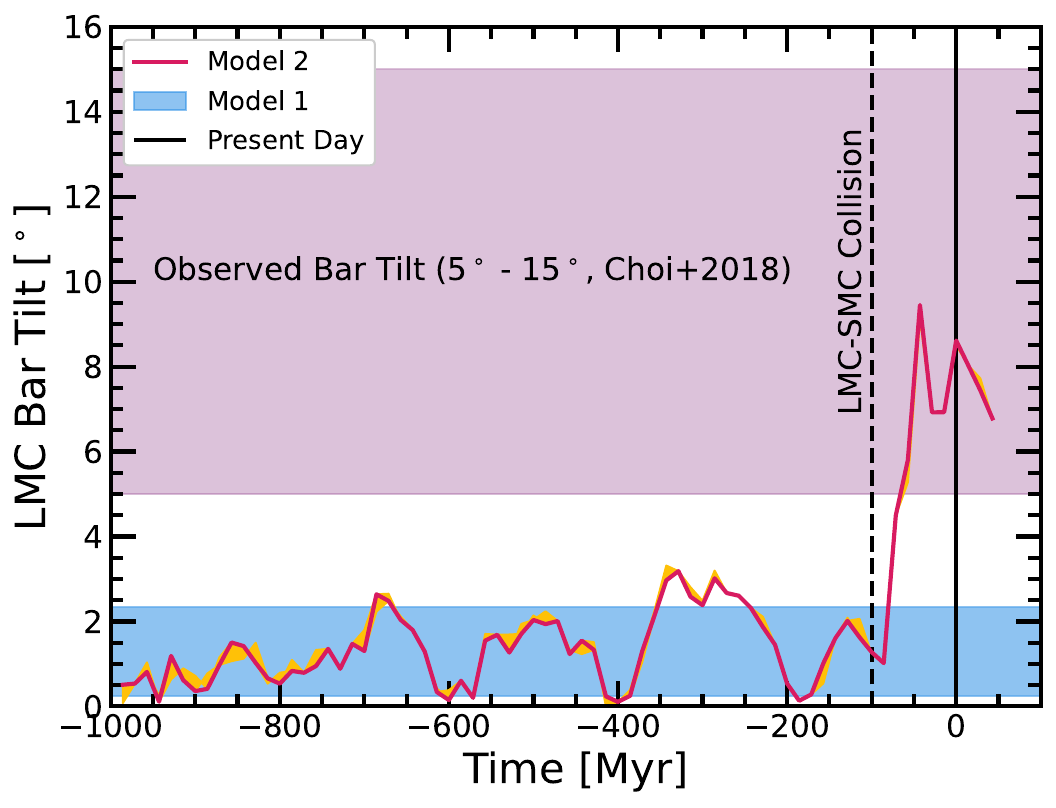}
    \caption{The simulated LMC bar's tilt as a function of time. The red solid line is the measured tilt in the Model 2 simulation and the shaded yellow band around it is the measurement error. The magenta shaded band depicts the observed bar tilt of $5^\circ - 15^\circ$ (C18). The vertical black dashed line is the LMC-SMC collision epoch and the vertical black solid line is the present day epoch in the simulation. The LMC bar tilts just as the SMC collides. At present day, in the simulation, the bar tilt ($\approx 8.6^\circ$) is consistent with the observations. The tilt remains consistent with observations for at least 100 Myr, persisting to the future for 50 Myr past the present day epoch. The blue shaded band denotes the $1-\sigma$ spread about the mean Model 1 bar tilt of $\approx 1^\circ$. Thus, in Model 1 (where the Clouds do not collide), the bar does not tilt sufficiently to explain the observations. 
    }
    \label{fig:tilt_time}
\end{figure}

In this section we study the vertical displacement (tilt) of the LMC bar with respect to the LMC disk plane, as simulated in B12 Model 2 and Model 1. 

We visualize the present-day simulated LMC bar's tilt with iso-density ellipse fits to the XZ projection of the stellar surface density distribution in the LMC bar frame of reference (defined in section \ref{sec:offset}). We choose the iso-density ellipse whose semi-major axis equals the bar radius as measured in section \ref{sec:offset}. This ellipse is referred to as the \enquote{edge-on bar ellipse}.

Figure \ref{fig:ellipse_tilt} shows the edge-on bar ellipse for three epochs in Model 2 - when the LMC and SMC are at the MW virial radius (MW infall epoch, left panel), when the LMC and SMC collide (LMC-SMC collision epoch, middle panel) and the present day epoch (right panel). A clear misalignment is visible between the semi-major axis of the edge-on bar ellipse and the horizontal disk plane in the present day snapshot, indicating a significant bar tilt. The misalignment between the edge-on bar ellipse and the horizontal disk plane is less in the LMC-SMC collision epoch and is negligible in the MW infall epoch. Thus, the LMC's bar develops a strong tilt with respect to the horizontal plane post SMC collision. The negligible tilt seen in the MW infall epoch of Model 2 is also representative of the tilt in the Model 1 simulation (where the Clouds do not collide).

Next, we quantify the bar tilt in the simulations as a function of time. The bar tilt corresponds to the position angle of the edge-on bar ellipse (Figure \ref{fig:ellipse_tilt}) in the XZ plane. Following the approach in section \ref{sec:offset}, we estimate the measurement error in the bar tilt by computing the position angles of the edge-on bar ellipses whose semi-major axis is one softening length different from the bar radius.

Figure \ref{fig:tilt_time} shows the absolute value of the LMC bar's tilt as a function of time in Model 2. We mark the epoch of LMC-SMC collision as a dashed black line, and the present day epoch as a solid black line. The bar tilt increases significantly just as the SMC collides with the LMC, indicating a strong causal connection between the SMC collision and the LMC bar's tilt.

The observed LMC bar tilt of $5^\circ - 15^\circ$ \citet[hereafter C18]{Choi2018a} is shown as a pink shaded band in Figure \ref{fig:tilt_time}. The tilt at present day in the simulation ($\approx 8.6^\circ$) is consistent with the observed range. The tilt remains consistent with observations for at least 100 Myr. It has not significantly decayed even 50 Myr into the future. Hence, the tilt will be consistent with observations even if the collision occurred 150 - 200 Myr ago (timing inferred from the bar offset, section \ref{sec:offset}).

For comparison, we also compute the bar tilt as a function of time in the Model 1 simulation. The bar tilt in Model 1 remains small ($<2^\circ$) throughout, and fluctuates between $0^\circ$ and $2^\circ$ with a mean of $\approx 1^\circ$ and a standard deviation of $\approx 1^\circ$. Thus, Model 1 is not able to reproduce the tilt seen in Model 2 and in observations at the level of $6-\sigma$ and $2-\sigma$, respectively.

\subsection{Explaining the LMC's Tilted-Ring Structure}

\begin{figure*}
    \centering
    \includegraphics[height = 0.3\textwidth, width = 0.3\textwidth]{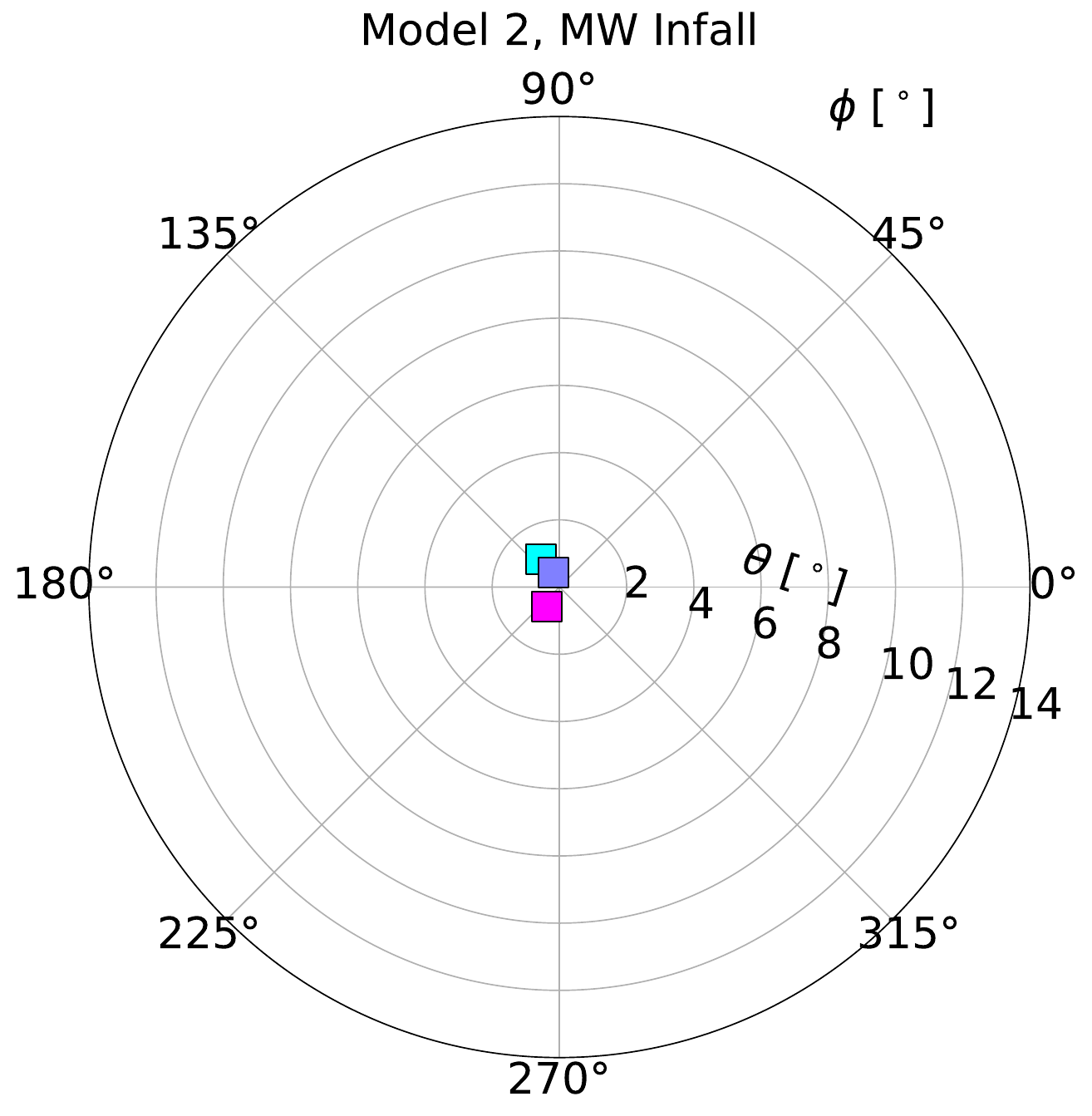}
    \includegraphics[height = 0.3\textwidth, width = 0.3\textwidth]{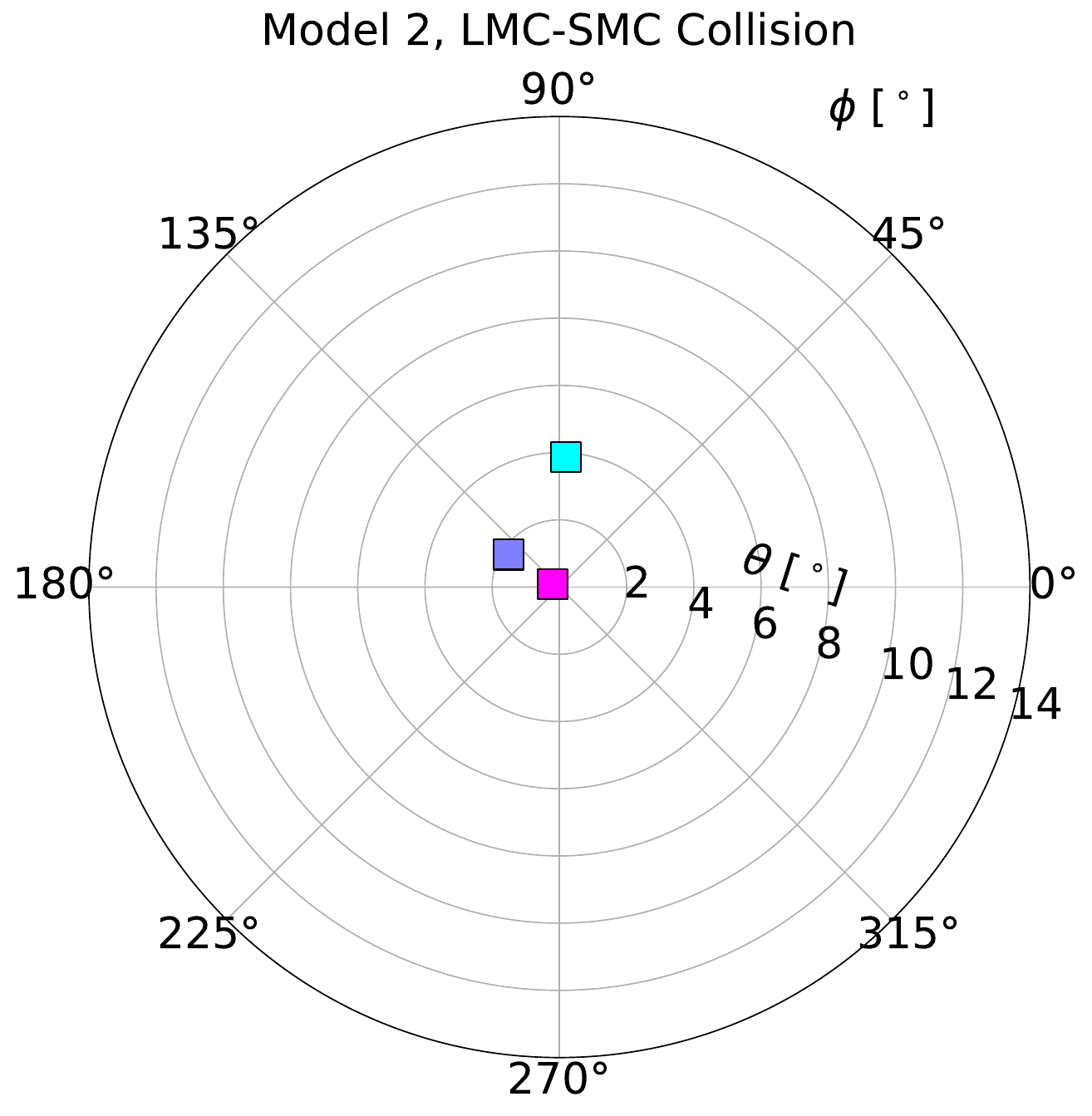}
    \includegraphics[height = 0.3\textwidth, width = 0.35\textwidth]{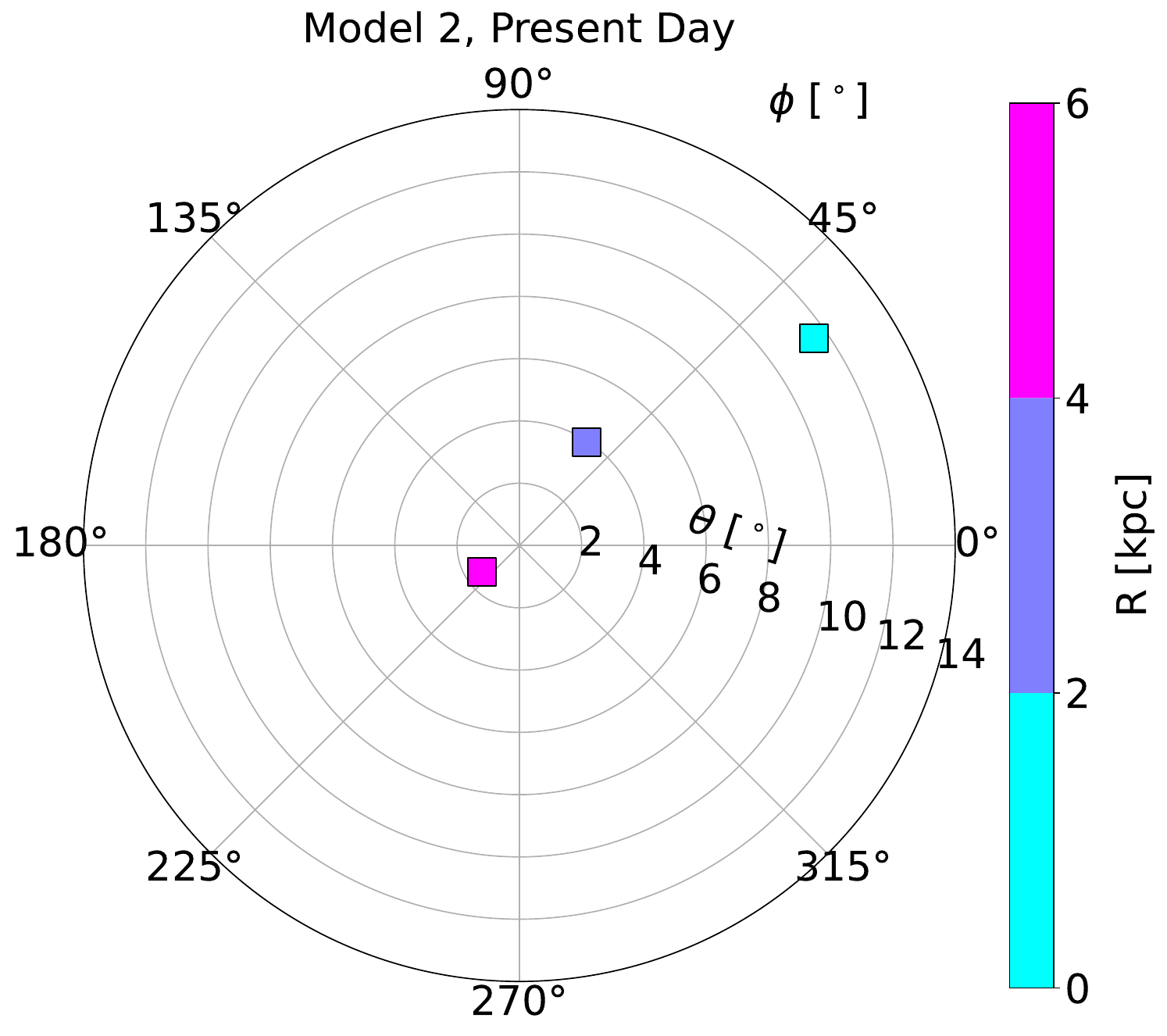}
    
    \caption{Briggs plots for the simulated LMC's stellar disk at three epochs, illustrating the angle between the average angular momentum vectors of specific radial annuli ($\vect{L_i}, \:\: i = 0, 1, 2$) and the total angular momentum vector of the disk within 10 kpc (aligned with the Z-axis). The disk is binned in three annuli - 0 to 2 kpc (inner disk dominated by the bar, cyan square marker), 2 - 4 kpc (transition region between the bar and the outer disk, blue square marker) and 4 - 6 kpc (outer disk, magenta square marker). The $\theta$ axis of each panel is the angle between $\vect{L_i}$ and the Z-axis. The $\phi$ axis of each panel is the azimuth of $\vect{L_i}$ projected in the XY plane. {\em Left panel}: 1 Gyr ago, when the LMC and SMC infall towards the MW. $\vect{L_i}$ of each annulus is well-aligned with the Z-axis, indicating a negligible bar tilt. {\em Middle panel}: 100 Myr ago, when the LMC and SMC collide. The angular momentum vector of the inner disk starts to become misaligned with the Z-axis, indicating that the bar is in the process of tilting. {\em Right Panel}: The present day. The inner disk is significantly misaligned with the Z-axis and the outer disk (by $\approx 10^\circ$), indicating a strong bar tilt. Further, the misalignment of the transition region with the Z-axis ($\approx 4^\circ$) is in between that of the inner disk and outer disk. The Briggs plot at present day is consistent with the observed tilted-ring structure of the LMC \citep{Arranz2025}, which is naturally explained by the LMC-SMC collision.}
    \label{fig:briggs}
\end{figure*}

\begin{figure}
    \centering
    \includegraphics[width=\columnwidth]{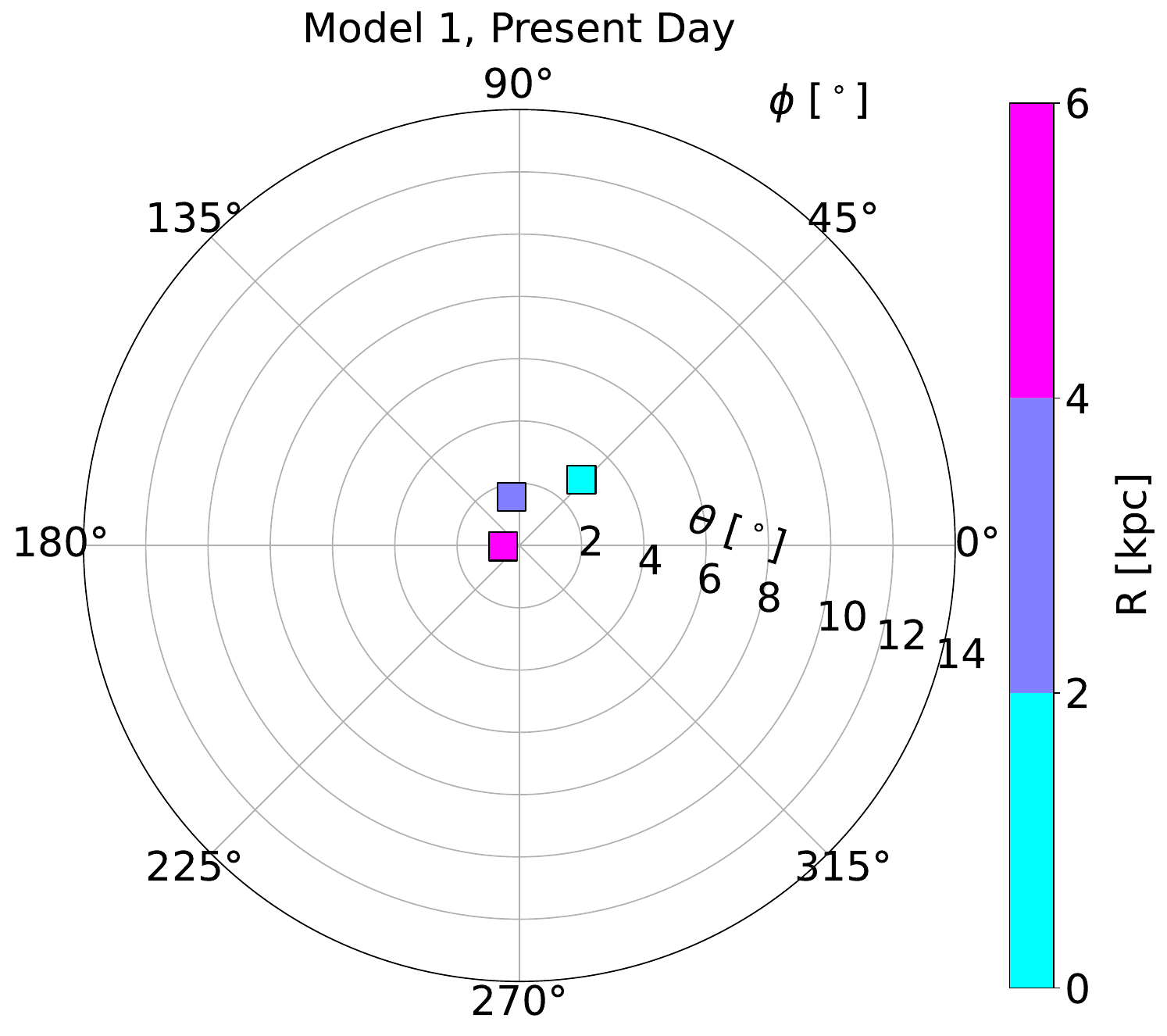}
    \caption{Same as Figure \ref{fig:briggs}, but for the present day snapshot of the Model 1 simulation, where the LMC and SMC do not collide. The angular momentum vector of the inner LMC disk (dominated by the bar, 0 - 2 kpc) remains roughly aligned with the outer LMC disk (4 - 6 kpc). This is inconsistent with observations (JA25).}
    \label{fig:briggs_m1}
\end{figure}

\citet[hereafter JA25]{Arranz2025} built a kinematic model of the LMC's disk using proper motions of LMC stars from Gaia-DR3 combined with line of sight velocities from Gaia-DR3, SDSS-IV and SDSS-V. They find that the LMC's kinematics is best explained if the inner disk dominated by the bar ($R < 2$ kpc) is inclined with respect to the outer disk (4 - 6 kpc) by $\approx 10^\circ$. Further, the transition region (2 - 4 kpc) has an inclination value in between the inner disk and outer disk. Such a disk morphology is best represented by a tilted-ring structure \citep{Rogstad1974}, wherein the stellar distribution and kinematics of a galaxy can be described as a set of concentric but mutually inclined, rotating rings (see Figure 13 of JA25 for a visualization). The LMC's tilted-ring structure is also consistent with the morphology inferred by C18 using distances to red clump stars, where the inclination of the inner ring (0 - 2 kpc) corresponds to the bar tilt.

In this section, we explore whether the LMC-SMC collision can explain the tilted ring structure of the LMC out to 6 kpc. We verify this by constructing the angular momentum profile of the simulated LMC's disk, also known as the Briggs plots \citep{Briggs1990}. Briggs plots have been used before to represent warps in the disk \citep[e.g.][]{Khachaturyants2022}. Below, we outline the construction of the Briggs plots.

We work in the LMC bar frame of reference, where the direction of the total angular momentum vector of the disk within 10 kpc ($\vect{\hat{L}}$) is aligned with the Z-axis (but the bar itself can still be tilted with the disk plane). We bin the disk in three radial annuli - 0 kpc to 2 kpc (the inner disk, which is dominated by the stellar bar), 2 kpc - 4 kpc (the transition region between the stellar bar and the outer disk), and 4 kpc - 6 kpc (which corresponds to the outer disk). We compute the average angular momentum of star particles in each radial annuli, and refer to this quantity as $\vect{L_i}$ for the $i^{th}$ annulus. Then, we compute the following angles:
\begin{equation}
\theta_i = \arccos{\left(\vect{\hat{L}}\cdot\vect{\hat{L_i}}\right)}
\end{equation}
\begin{equation}
\phi_i = \arccos{\left(\frac{\vect{L_{i,y}}}{\vect{L_{i,x}}}\right)}
\end{equation}
\noindent where, $\theta_i$ is the angle between the angular momentum vector of the $i^{th}$ annulus and the Z-axis, $\phi_i$ is the azimuth of the angular momentum vector of the $i^{th}$ annulus when projected onto the disk plane and measured with respect to the X-axis in an anti-clockwise sense. 

The Briggs plot is constructed as a polar projection with $\theta_i$ along the radial axis and $\phi_i$ along the azimuth axis. If any radial annulus has a different direction of angular momentum from the Z-axis, it will be represented as a point with a non-zero ($\theta_i$, $\phi_i$) in the Briggs plot. We estimate a typical error on the points constituting the Briggs plot with the angle subtended by the simulation softening length towards the disk center at a given radius. By taking the midpoints of the three radial annuli, the Root Mean Square error across the three annuli turns out to be $\approx2^\circ$. We refer to this error as $\sigma_{\rm{Briggs}}$. 

Figure \ref{fig:briggs} shows the Briggs plots for three epochs in the B12 Model 2 simulation, as defined in Figure \ref{fig:ellip_offset}: 1) when the LMC and SMC are at the MW virial radius (MW infall epoch, left panel); 2) when the LMC and SMC collide (LMC-SMC collision epoch, middle panel); and 3) the present day epoch (right panel). At the MW infall epoch, almost all the radial annuli are clustered near the origin, which indicates that the disk has a uniform direction of angular momentum throughout, consistent with a negligible bar tilt and coherent disk planarity. 

At the LMC-SMC collision epoch, the inner disk (0-2 kpc), which is dominated by the bar, has a different $\theta_i$ value than the larger annuli (2-6 kpc), whose $\theta_i$ values are close to the origin. This indicates that the angular momentum vector of the inner disk/bar starts to get misaligned with that of the outer disk at the time of collision. 

At the present day epoch, 100 Myr after collision (Figure \ref{fig:orbit_vis} and the right panel of Figure \ref{fig:ellipse_tilt}), the angular momentum vector of the inner disk is strongly misaligned (by $\approx 10^\circ$) from the Z-axis, indicating a strong bar tilt. The outer disk (4 - 6 kpc) is still aligned with the Z-axis within $\sigma_{Briggs}$. Further, the misalignment of the transition region (2 - 4 kpc) with the Z-axis is in between the inner disk and outer disk. The Briggs plot at present day reflects the tilted-ring structure of the simulated LMC, which is naturally explained by a collision with the SMC.

For comparison, in Figure \ref{fig:briggs_m1}, we construct the Briggs plot for the present day snapshot of the Model 1 simulation (which corresponds to the left panel of Figure \ref{fig:ellipse_tilt}). In the Model 1 simulation, the angular momentum vectors of all the radial annuli are aligned with the Z-axis within $\sigma_{Briggs}$ at all times, which is inconsistent with observations.
\\
\\
To the best of our knowledge, this is the first time tilted bars have been studied in such detail in a numerical simulation. \cite{Collier2023} studied the evolution of a simulated galactic bar in a counter-rotating dark matter halo and suggested that the bar can develop a tilt due to resonant exchanges of angular momentum with dark matter. However, they did not characterize the bar tilt and its evolution in detail. In section \ref{sec:toy_model}, we show that the torques applied by the SMC on the LMC's bar during the collision are the likely cause of the bar tilt. Further, using the torque calculations, we demonstrate how the SMC's dark matter profile can be constrained using the LMC's observed bar tilt.

\section{The LMC Bar's Pattern Speed} \label{sec:ps}

\begin{figure}
    \centering
    \includegraphics[width=\columnwidth]{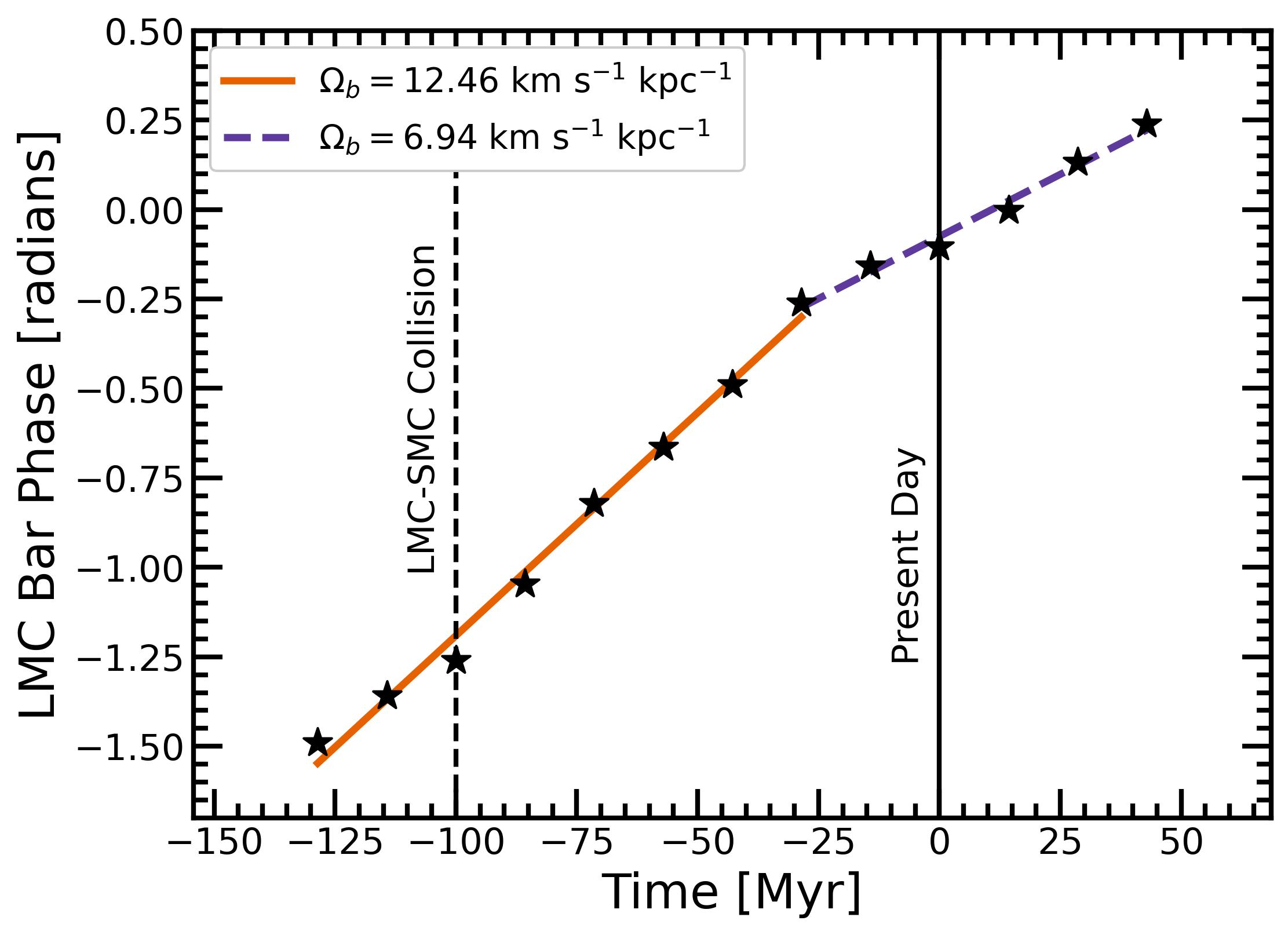}
    \caption{Evolution of the LMC bar pattern speed in the B12 Model 2 simulation. The m = 2 phase angle of the bar is plotted as a function of time (black stars). The epoch of the LMC-SMC collision (black dashed line), and the present day ($t = 0$, black solid line) are marked. The slope of the linear fit to the bar phase vs. time yields the pattern speed of the bar. The evolution of the phase of the LMC's bar is described by two pattern speeds: 1) $\Omega_b = 12.46$ km s$^{-1}$ kpc$^{-1}$ prior to and during the LMC-SMC collision (solid orange line);  and 2) $\Omega_b = 6.94$ km s$^{-1}$ kpc$^{-1}$ at present day/after the collision (purple dashed line). The change in the pattern speed is larger than the scatter in the time evolution of the pattern speed owing to secular evolution ($\approx 2.4$ km s$^{-1}$ kpc$^{-1}$). Thus, a recent collision with the SMC can slow down the pattern speed of the LMC's bar by a factor of 2. Hence, caution must be exercised in the interpretation of the LMC bar's formation history with the observed pattern speed values.}
    \label{fig:ps_model2}
\end{figure}

Following literature \citep[e.g.][]{Debattista1998, Silva2023, Valluri2016}, we compute the pattern speed of the simulated LMC bar using the $m = 2$ phase angle ($\Phi_{bar}$), which was already computed for the B12 simulations in section \ref{sec:offset}. The pattern speed is then given by:
\begin{equation} \label{eq:bar_ps}
    \Omega_{b} = \frac{d\Phi_{bar}}{dt}
\end{equation} 
We fit a line to the evolution of $\Phi_{bar}$ as a function of time, and compute the slope of the best fit line, yielding a best fit value for $\Omega_b$. The Nyquist frequency at the time cadence of the snapshots (14 Myr) is $449$ Gyr$^{-1}$ (or 439 km s$^{-1}$ kpc$^{-1}$), which is much larger than the expected pattern speed of a galactic bar ($<\sim50$ km s$^{-1}$ kpc$^{-1}$). Hence, the time cadence of the snapshots is sufficient to sample the simulated LMC bar phase. 

Figure \ref{fig:ps_model2} shows the resulting time evolution of bar phase for the B12 Model 2 simulation. The evolution of the bar phase is best described by two pattern speeds: 1) $\Omega_b = 12.46$ km s$^{-1}$ kpc$^{-1}$ at (and prior to) the LMC-SMC collision epoch;  and 2) $\Omega_b = 6.94$ km s$^{-1}$ kpc$^{-1}$ at the present day epoch.  This result indicates that the LMC-SMC collision can change the pattern speed of the LMC's bar by almost a factor of 2. The change in pattern speed is not instantaneous. It takes $\approx50$ Myr for it to change.

To confirm that the change in bar pattern speed is not a result of secular evolution, MW tides, or weak SMC tides, we measure the bar pattern speed in Model 1 (where the Clouds do not collide). We apply finite differencing to the $m = 2$ phase of consecutive snapshots. We clip pattern speeds that are outside $2-\sigma$ on either side of the mean to remove the extreme outliers resulting from noise in finite differencing. We find the average Model 1 pattern speed to be $12.92$ km s$^{-1}$ kpc$^{-1}$ with a scatter of $\approx 2.4$ km s$^{-1}$ kpc$^{-1}$ over the MW infall duration. The change in pattern speed of Model 2 post collision is more than twice this scatter, which indicates that the change in pattern speed is driven by the SMC's collision.

This result is consistent with theoretical expectations. \cite{Gerin1990} performed N-body simulations of a disk galaxy interacting with a spherical companion (mass ratio of 0.5). They considered both equatorial (retro-grade and pro-grade to the disk) and vertical encounters. The scale length of the primary galaxy disk was 4 kpc, and the smallest pericenter approach of the secondary was 5 kpc. They found that bars can both slow down and speed-up depending on the direction of torques applied by the satellite (prograde vs. retrograde to the disk). \cite{Sundin1991, Sundin1993} performed an upgraded version of the \cite{Gerin1990} simulations wherein they considered more mass ratios and tried to setup a more stable primary disk with lesser variation of the bar pattern speed in isolation. They found that the direction of torques alone is insufficient to explain whether the bar slows down or speeds up during the interaction. They suggest that the mass ratio of the host:satellite is also important in addition to resonant exchanges of angular momentum between the bar and satellite. 

APB97 (see section \ref{sec:offset} for their simulation details) found that the bar pattern speed dropped by a factor of 2-3 just after the satellite impact in almost all of their simulations. Furthermore, B03 found that for impacts that happened $\approx 150$ Myr ago, the bar pattern speed reduced by a factor of $1.5$. The reduction in bar pattern speeds observed by APB97 and B03 is consistent with our findings.

In the KRATOS simulations, \cite{Arranz2025b} find that the LMC bar pattern speed can slow down significantly due to the SMC's interactions. In some interacting configurations, they find that the LMC bar can even momentarily stop rotating.

Since the SMC's collision can change the pattern speed of a pre-existing LMC bar by a factor of a few, observations of the LMC bar pattern speed should be interpreted with caution. The observed pattern speed values cannot directly be used to place the LMC's bar in context with other barred galaxies of the local universe or for understanding bar-driven secular evolution in the LMC.

Further, methods for measuring the pattern speed in observations like the Tremaine-Weinberg method \citep{TW1984}, the Dehnen method \citep{Dehnen2023}, the bi-symmetric velocity method \citep{Luri2021}, or inferences from Schwarzschild dynamical modeling \citep{Schwarzschild1979}, rely on the assumption of a stable bar that has not suffered from significant external perturbations. However, in this work we have shown that the LMC's bar is likely in a high state of dis-equilibrium due to a LMC-SMC collision. Likely, less than one bar dynamical time (estimated in section \ref{sec:toy_model}) has passed since the collision. Thus, the validity of the aforementioned methods in such a highly disequilibrium scenario needs to be investigated further. Given the high degree of disequilibrium, it is not surprising that different methods give completely different values of the bar pattern speed when applied to the LMC's observational data \citep[e.g.][]{Arranz2024}.

A detailed analysis of the exact dynamical mechanism(s) through which the SMC affects the LMC bar's pattern speed will be the subject of a future work. Post SMC collision, the simulated LMC bar's geometry changes significantly. In addition to the bar developing an offset and a tilt, the bar also shortens (in radius) by $\approx 0.5$ kpc within the time between the collision and the present day. Hence, the evolution of the bar pattern speed post-collision is likely dictated by a combination of several dynamical effects, including the SMC's direct torques on the bar and the collision-induced changes in the bar's geometry.

\section{Further Analysis and Discussion}  \label{sec:discussion}

We find that a recent collision between the LMC and SMC naturally explains the LMC bar's observed offset and tilt. The collision can also significantly affect the LMC bar's pattern speed.  In this section we: 1) provide a framework to quantitatively compare the LMC's gas distribution in different LMC-SMC interaction scenarios (section \ref{sec:gas_bar}); 2) discuss how the properties of the LMC's bar can constrain the SMC's dark matter profile (section \ref{sec:toy_model}); and discuss the 3) limitations of our work and future directions of study (section \ref{sec:limitations}). 

\subsection{A Framework to Study the Simulated Central Gas Distributions}
\label{sec:gas_bar}

\begin{figure*}
    \centering
    \includegraphics[height = 0.3\textwidth, width = 0.3\textwidth]{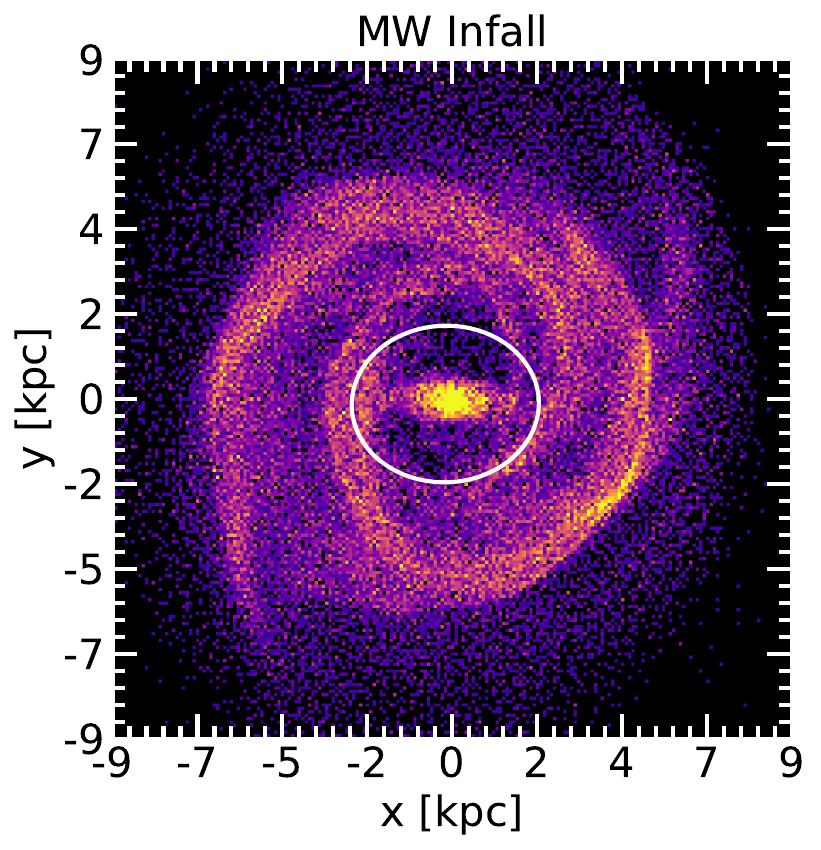}
    \includegraphics[height = 0.3\textwidth, width = 0.3\textwidth]{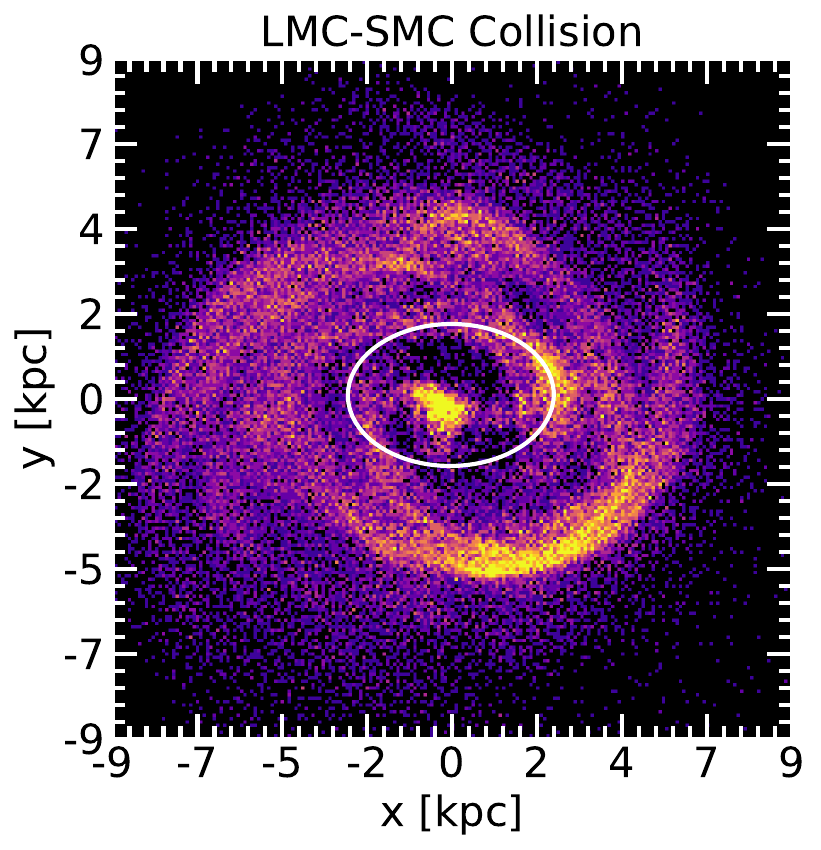}
    \includegraphics[height = 0.3\textwidth, width = 0.35\textwidth]{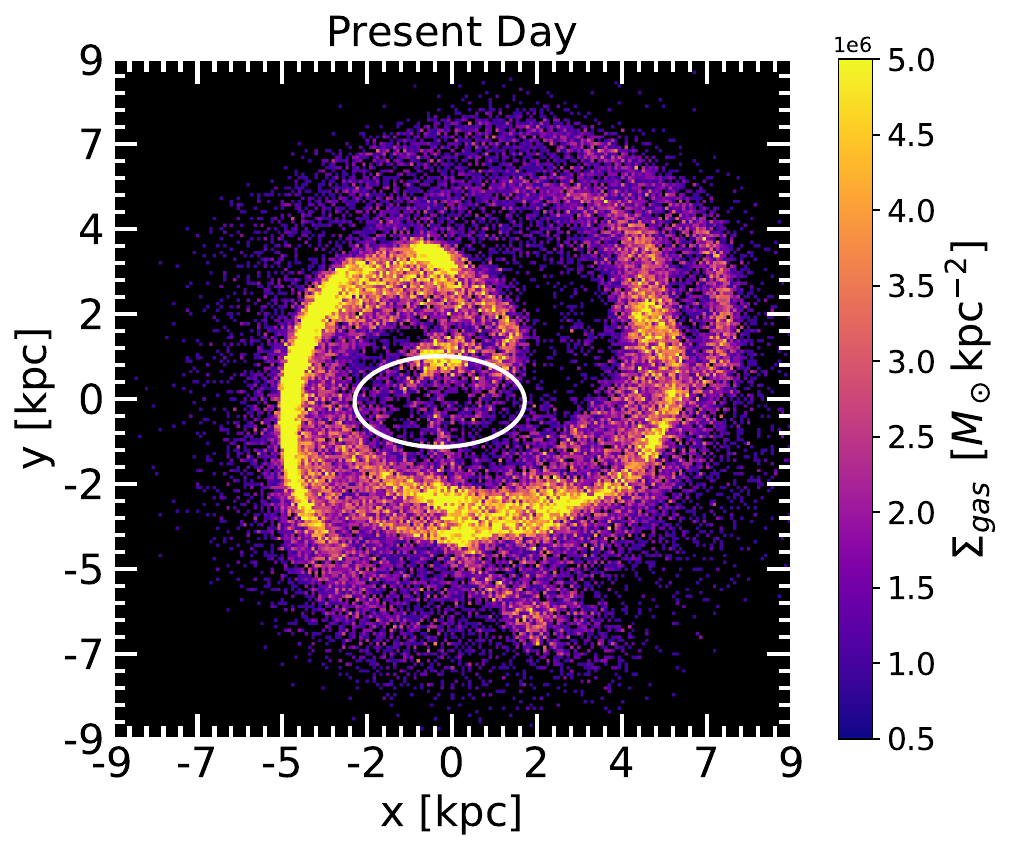}
    \caption{Surface gas density distribution ($\Sigma_{gas}$) in the simulated LMC disk of B12 Model 2, plotted face-on
    in the LMC bar frame (defined in section \ref{sec:offset}) at three different epochs. {\em Left panel}: 1 Gyr ago, when the LMC and SMC infall into the MW. {\em Middle panel}: 100 Myr ago, when the LMC and SMC collide. {\em Right panel}: the present day. The white isodensity ellipse indicates the stellar bar, as shown in Figure \ref{fig:ellip_offset}. At infall, gas forms an elongated structure along the semi-major axis of the stellar bar. When the LMC and SMC collide, the elongated gas feature starts to disrupt. At the present day, the density of gas in the bar region has dropped by a factor of $\sim$10. The prominent elongated gas feature at the MW infall epoch is also representative of the LMC's gas distribution in the Model 1 simulation at present day (where the Clouds do not collide).}
    \label{fig:gas_maps}
\end{figure*}

\begin{figure}
    \centering
    \includegraphics[width=\columnwidth]{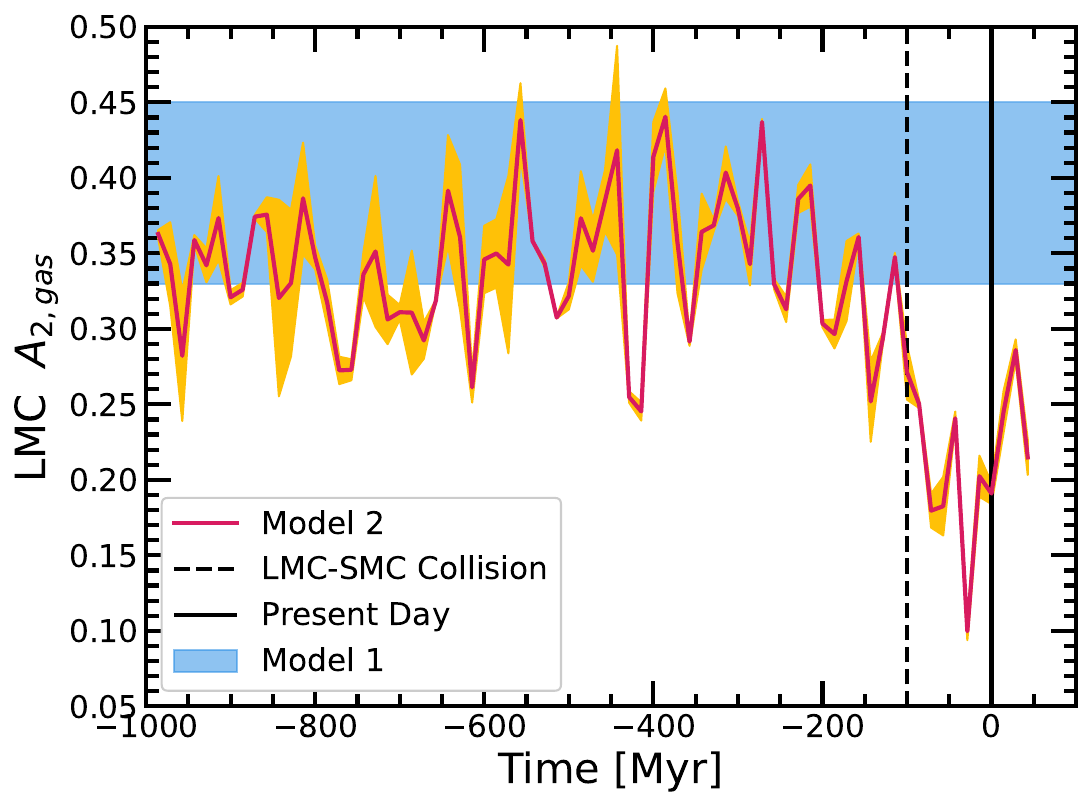}
    \caption{The normalized m = 2 Fourier amplitude of the gas surface density distribution ($A_{2, gas}$) in the B12 Model 2 LMC's central region as a function of time. $A_{2, gas}$ quantifies the strength of an elongated structure in gas. The red solid line indicates the measurement for the Model 2 simulation and the shaded yellow band around it denotes the measurement error. The vertical black dashed line marks the LMC-SMC collision epoch and the vertical black solid line marks the present day epoch. We find that the strength of an elongated feature in gas decreases significantly when the SMC collides with the LMC. The blue shaded band denotes the $1-\sigma$ spread on either side of the mean $A_{2, gas}$ (= 0.39) in the Model 1 simulation calculated over 600 Myr prior to the present day. In Model 1, where the LMC and SMC do not collide, gas maintains a prominent presence in the LMC's bar at all times.}
    \label{fig:gas_time}
\end{figure}

In B12, it has been shown that the LMC's bar region in Model 2 is devoid of gas (see Figure 13 of B12). Whereas, in the Model 1 simulation, gas maintains a prominent presence in the bar region (see Figure 12 of B12). In this section, we present a framework to study the evolution of gas in the simulated LMC's bar region to enable the utilization of the gas distribution as another complementary probe of the LMC-SMC interaction history. In particular, we show that the machinery we have developed for understanding the LMC's stellar bar can also be utilized to understand its central gas distribution. This presented framework is generalizable to more advanced hydrodynamic simulations.

Figure \ref{fig:gas_maps} shows the surface density maps of the LMC's gas disk for three epochs - when the LMC and SMC are at the MW virial radius (MW infall epoch, left panel), when the LMC and SMC collide (LMC-SMC collision epoch, middle panel) and the present day epoch (right panel). We also overlay the isodensity ellipse corresponding to the stellar density at the end of the stellar bar  (refer to Figure \ref{fig:ellip_offset}).

Prior to the SMC collision, gas forms an elongated structure along the semi-major axis of the stellar bar. This structure is created because of gas inflows driven by the bar. These inflows have been characterized before in hydrodynamic simulations \citep[e.g.][]{Weiner1999, Berentzen2003, Fanali2015, Sormani2015}. At the LMC-SMC collision epoch, the elongated structure in gas starts to disrupt. At present day, the density of gas in the central regions (within $\sim 2$ kpc) has decreased by more than an order of magnitude as compared to the pre-collision epochs. A central elongated structure in gas is non-existent at present day. The prominent gas feature at the MW infall epoch (left panel) is also representative of the gas distribution in the Model 1 simulation (where the Clouds do not collide).

We quantify the weakening of the elongated gas feature with the Fourier decomposition method described in section \ref{sec:offset}, applied to the gas surface density distribution. Fourier amplitudes have been used before to describe non-axisymmetric perturbations in gas disks \citep[e.g.][]{Liang2024}.

First, we align the snapshots to the LMC bar frame (section \ref{sec:offset}). Then, we compute the center of mass of the gas distribution with the same iterative shrinking sphere approach used to compute the stellar center of mass (section \ref{sec:com}). We use the gas center of mass for centering the spatial Fourier decomposition of the gas surface density. Similar to section \ref{sec:offset}, we determine the radius (semi-major axis) of the elongated gas feature by observing the variation in the $m = 2$ phase of the gas distribution as a function of the radial coordinate. We call this radius $R_{m = 2, gas}$. The average $R_{m = 2, gas}$ across the simulation snapshots prior to collision in Model 2 is $\langle R_{m = 2, gas}\rangle \approx 1$ kpc (a factor of 3 smaller than the mean stellar bar radius over the same timescale). 

Next, we compute the m = 2 amplitude of the gas distribution within a radial aperture of $\langle R_{m = 2, gas}\rangle$ around the gas center of mass. We refer to this amplitude as $A_{2, gas}$. We vary the radius of the aperture by one softening length to obtain an error estimate on $A_{2, gas}$.

Figure \ref{fig:gas_time} shows $A_{2, gas}$ as a function of time for the Model 2 simulation as a red solid line. The yellow shaded band on the red solid line is the measurement error. $A_{2, gas}$ significantly decreases when the SMC collides with the LMC. This is consistent with Figure \ref{fig:gas_maps}.

We similarly compute $A_{2, gas}$ for Model 1 (where the LMC and SMC do not collide). In Model 1, $A_{2, gas}$ decreases from $\approx 0.5$ to $\approx 0.4$ over for a period of 400 Myr after MW infall, which is possibly a result of secular evolution/MW tides. However, $A_{2, gas}$ remains roughly constant for the remainder of the simulation time, indicating that the gas inflow along the bar has stabilized. After this stabilization, the mean and standard deviation of $A_{2, gas}$ in Model 1 is $0.39$ and $0.06$ respectively. Gas maintains a prominent presence in the LMC bar region in the Model 1 simulation at all times, which is inconsistent with the evolution seen in Model 2.   

The framework we have developed in this section provides a quantitative means to compare gas distribution in different LMC-SMC interaction scenarios. However, with the B12 simulations, it is difficult to directly compare the gas distribution to observations, primarily because Supernova feedback is not implemented. Feedback can significantly affect the gas distribution in a galactic disk. It can remove gas over short timescales by injecting energy into the Interstellar medium. It can also support gaseous inflows over long timescales by enriching the Interstellar medium with metals and enhancing radiative cooling.

Hence, explaining the lack of gas in the observed LMC's central regions is a pressing question, and needs to be explored further with more advanced hydrodynamic prescriptions, which we will pursue in the future. LMC-SMC interactions will definitely play a role in shaping the gas distribution of the Clouds, however, the relative influence of secular and interaction driven processes needs to be studied further. Future work also involves studying the role of an offset and tilted stellar bar in shaping the gas distribution. Understanding these aspects is crucial to interpret the stellar populations in the LMC bar region, wherein it is found that the bar is dominated by old stars (age $>$ 1 Gyr) \citep{Harris2009, ElYoussoufi2019, Mazzi2021, Luri2021}.

\subsection{A Semi-Analytic Model to Understand the LMC Bar's Tilt and Constrain the SMC's Mass Profile} \label{sec:toy_model}

In section \ref{sec:tilt}, we demonstrated a causal connection between the SMC's collision and the LMC bar's tilt. In section \ref{sec:intro}, we motivated how the unusual properties of the primary's bar can be used to constrain the mass profile of the satellite galaxy in interacting systems. In this section, we show that the LMC bar's tilt encodes information about the SMC's pre-collision dark matter profile. Decoding this information requires us to model the torques applied by the SMC on the LMC's bar.

We develop a simplified and instructive semi-analytical model (SAM) of the torques applied by the SMC on the LMC's bar. The model is semi-analytical in the sense that the SMC is assumed to follow the same orbit as the B12 Model 2 simulation, but the SMC's gravitational torques are computed analytically. Further, the torques are assumed to be impulsive, which means the timescale over which significant torques are applied must be significantly smaller compared to the relevant dynamical timescale of the LMC's bar. This is indeed true in the Model 2 scenario. The dynamical timescale of the LMC's bar ($\sim \frac{2\pi}{\Omega_b} \approx 600$ Myr) is much larger than the timescale of the collision ($\sim 50$ Myr). The significant difference in the timescales means the impulse approximation can be used to calculate the effect of the SMC's torques on the LMC's bar.

Let's set up the problem. We orient the snapshots in the LMC bar frame (see section \ref{sec:offset}). Let the SMC's position vector relative to the LMC's center, at some instant of time $t$, be:
\begin{equation} \label{eq:smc_pos_vec}
    \vect{r_{SMC}(t)} = [x_{SMC}(t), \: y_{SMC}(t), \: z_{SMC}(t)]
\end{equation}
In this simplified approach, the SMC is assumed to be a point mass perturber. This is a reasonable assumption since the dominant contribution of the torques will be from the SMC's mass present in a sphere of radius comparable to the impact parameter of the collision ($\approx2$) kpc. This mass will be referred to as $M_{SMC}(< 2 \: \rm{kpc})$.  

\begin{figure}
    \centering
    \includegraphics[width=\columnwidth]{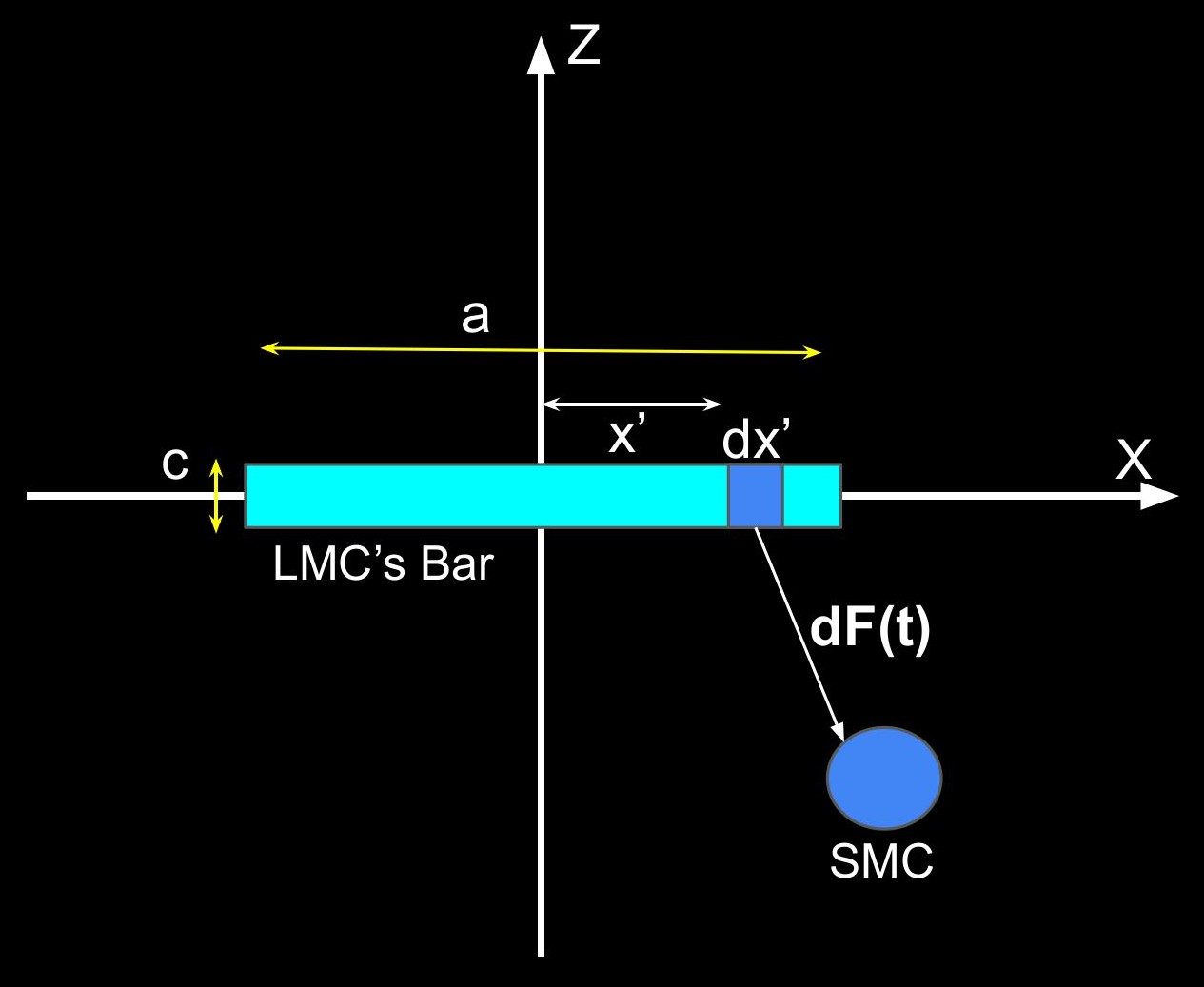}
    \caption{A schematic describing a semi-analytic model to compute the gravitational torques of the SMC on the LMC's bar. The LMC's bar is modeled as a solid cuboid with length $a$, width $b$ and height $c$. Coordinates are oriented such that the disk is aligned with the XY plane and the bar is aligned with the X-axis in each snapshot. The schematic depicts the X-Z projection where the disk is viewed edge-on. The SMC is treated as a point mass. To compute the torque, we consider a differential length element of the bar $(dx')$ and calculate the force $\mathbf{dF}$ that the SMC exerts on that length element at a time $t$. We calculate the torque on the length element, and integrate it across the length of the bar to obtain the total torque on the bar at a time $t$. Then, we integrate in time to obtain the impulse of the SMC's torque and compute the resulting tilt of the LMC's bar due to the SMC's impulse.}
    \label{fig:toy_model}
\end{figure}

\begin{figure*}
    \centering
    \includegraphics[width = 0.45\textwidth, height = 0.35\textwidth]{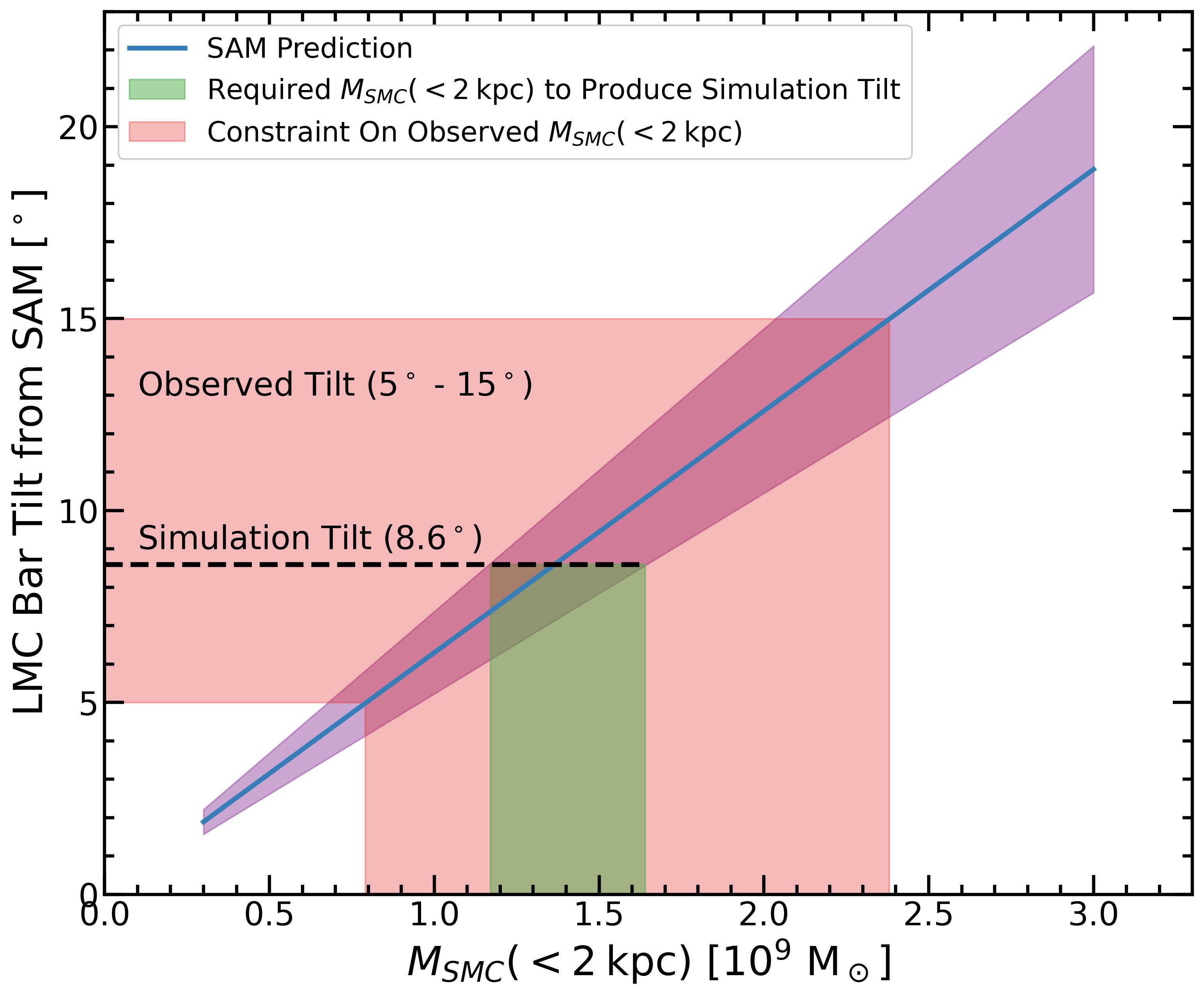}
    \includegraphics[width = 0.45\textwidth, height = 0.35\textwidth]{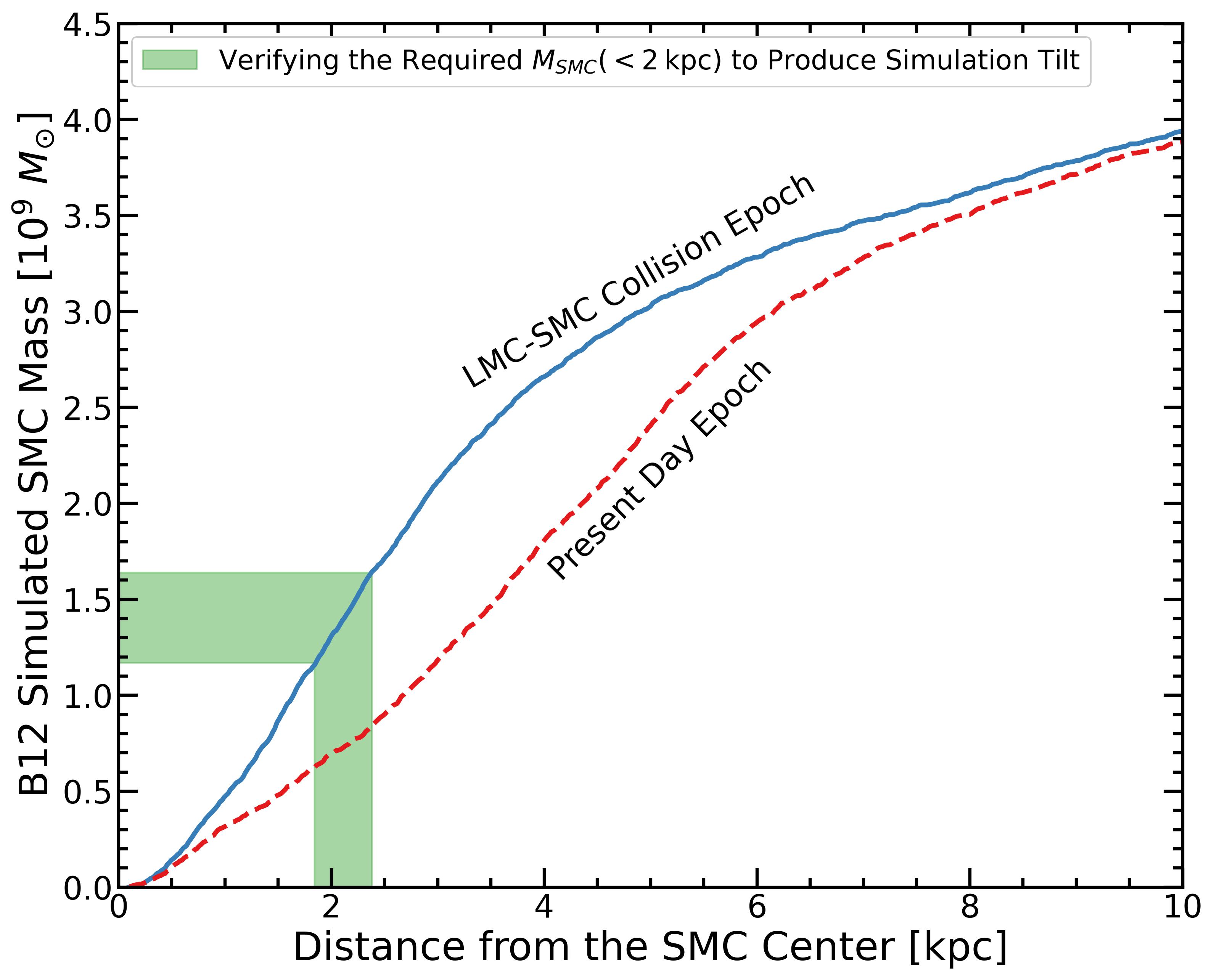}
    \caption{The framework to constrain the total mass of the SMC in its inner 2 kpc ($M_{SMC}(< 2 \: \rm{kpc})$) using a semi-analytic model (SAM) for the SMC's torques. The left panel shows the bar tilt calculated from the SAM (equation \ref{eq:theta_t}) as a function of $M_{SMC}(< 2 \: \rm{kpc})$. The dark blue solid line is the SAM prediction, and the purple shaded band denotes the $3-\sigma$ error in the prediction. The SAM predicts that $M_{SMC}(< 2 \: \rm{kpc}) \approx (1.2 - 1.6) \times 10^9$ $M_\odot$ (green shaded band) to produce the bar tilt in B12 Model 2 ($\approx 8.6^\circ$;  Figure \ref{fig:tilt_time}). The right panel shows the Model 2 SMC's total mass profile (dark matter + gas + stars) as a function of distance computed at two epochs: the LMC-SMC collision (solid blue line), and the present day (dashed red line). As a result of the collision, the inner mass profile of the SMC becomes shallower. The SAM prediction of $M_{SMC}(< 2 \: \rm{kpc})$ to reproduce the Model 2 bar tilt (left panel) is consistent with the mass enclosed within 2 kpc from the SMC's center at the time of collision (but is inconsistent with the present day mass profile at the $5\sigma$ level). This exercise indicates the SAM reasonably represents the results of Model 2. Applying this model to the observed tilt of $5^\circ-15^\circ$, implies that, prior to the collision, $M_{SMC}(< 2 \: \rm{kpc}) = (0.8 - 2.4) \times 10^9$ $M_\odot$ (red shaded band in the left panel). This mass is significantly larger than the simulated gas $+$ stellar mass within 2 kpc at the time of the collision ($\approx3 \times 10^8$ $M_\odot$). The observed LMC bar tilt thus requires that the SMC is a dark matter dominated galaxy.}
    \label{fig:bar_tilt_model}
\end{figure*}

The force from the SMC on a small part of the LMC's bar is:
\begin{equation} \label{eq:diff_force}
    \vect{dF}(t) = \frac{G M_{SMC}(< 2 \: \rm{kpc}) \: dM_{bar}}{|[\vect{r_{SMC}(t)} - \vect{x}^\prime]|^3} [\vect{r_{SMC}(t)} - \vect{x}^\prime]
\end{equation}
\noindent where $\vect{dF}(t)$ is the differential force on a differential bar element of length $dx^\prime$ and stellar mass $dM_{bar}$ (refer to Figure \ref{fig:toy_model} for the problem setup).

We assume the bar has a constant linear mass density $\mu$:
\begin{equation} \label{eq:mu}
    \mu = M_{bar}/a
\end{equation}
\noindent where $a$ is the bar length, which is twice the radius of the bar. For Model 2, $a = 5.56$ kpc at the LMC-SMC collision epoch. $M_{bar}$ is the mass of the bar and $dM_{bar} = \mu \: dx^\prime$. We show later (equation \ref{eq:theta_t}) that to compute the bar tilt, the SMC's torque must be divided by the moment of inertia of the bar. Hence, the final tilt is not expected to be sensitive to the exact mass distribution within the bar.

We have, 
\begin{equation}
    \vect{dF}(t) = \frac{G M_{SMC}(< 2 \: \rm{kpc}) \: \mu \: dx^\prime}{|[\vect{r_{SMC}(t)} - \vect{x}^\prime]|^3} [\vect{r_{SMC}(t)} - \vect{x}^\prime]
\end{equation}
The torque on the bar segment $dx^\prime$ about the center of the bar is:
\begin{equation} \label{eq:dtorque}
    \vect{d\tau}(t) = \vect{x}^\prime \times \vect{dF}(t)
\end{equation}
Thus,
\begin{equation} \label{eq:dtorque_evaluated}
    \vect{d\tau} (t) = \frac{G M_{SMC}(< 2 \: \rm{kpc}) \: \mu dx^\prime}{|[\vect{r_{SMC}(t)} - \vect{x}^\prime]|^3} [\vect{x}^\prime \times \vect{r_{SMC}}(t)]
\end{equation}
The total torque on the bar due to the SMC is obtained by integrating the differential torque (equation \ref{eq:dtorque_evaluated}) over the extent of the bar:
\begin{equation} \label{eq:total_torque}
    \vect{\tau}(t) = \int_{-a/2}^{a/2} \frac{G M_{SMC}(< 2 \: \rm{kpc}) \: \mu dx^\prime}{|[\vect{r_{SMC}(t)} - \vect{x}^\prime]|^3} [\vect{x}^\prime \times \vect{r_{SMC}}(t)]
\end{equation}
Next we calculate the bar tilt. Bar tilt is a consequence of the SMC's torque about the Y-axis in the coordinate setup of Figure \ref{fig:toy_model}, and
\begin{equation} \label{eq:tau_y}
    \tau_{y} (t) = I_y \: \frac{d^2\eta}{dt^2},
\end{equation}
\noindent where $I_y$ is the moment of inertia of the bar about the Y-axis and $\eta$ is the tilt. To compute $I_y$, we model the bar as a solid cuboid, with dimensions $a$ (length), $b$ (width) and $c$ (height). A more sophisticated model would assume the bar to be a solid ellipsoid. However, the moment of inertia of a solid ellipsoid with axes lengths corresponding to the dimensions of a solid cuboid is very similar to the moment of inertia of a solid cuboid. Thus, the results are not sensitive to the assumed bar shape. Bar length ($a = 5.56$ kpc) has already been calculated for the collision epoch. The width, $b = 3.9$ kpc, is the minor axis of the iso-density ellipse corresponding to the Model 2 bar in the face-on projection at the collision epoch (see Figure \ref{fig:ellip_offset}, middle panel). The height, $c = 1.66$ kpc, is the minor axis of the iso-density ellipse corresponding to the Model 2 bar in the edge-on projection at the collision epoch (see Figure \ref{fig:ellipse_tilt}, middle panel).

The moment of inertia of the bar about the Y-axis is then:
\begin{equation} \label{eq:I_y}
    I_y = \frac{1}{12} M_{bar}  (a^2 + c^2).
\end{equation} 
The mass of the bar ($M_{bar}$) is computed as the total mass of the LMC stars present in the bar region, defined by the cuboid with dimensions $a$, $b$ and $c$. For the B12 Model 2 at the collision epoch, $M_{bar} = 10^9 \: M_\odot$.

Finally, the bar tilt is given by the solution of the differential equation (\ref{eq:tau_y}):
\begin{equation} \label{eq:theta_t}
    \eta(t) = \int_0^t \int_0^{t^\prime} \frac{\tau_y(t^{\prime\prime})}{I_y} dt^{\prime \prime} dt^{\prime} 
\end{equation}

Equation (\ref{eq:theta_t}) is integrated for a time interval where the impulse approximation is valid. That is, the time interval when the SMC's torques are significantly more dominant as compared to the restoring forces from the LMC disk and halo. This time interval is defined using the time derivative of the bar tilt (the \enquote{tilt speed}) after the collision. From Figure \ref{fig:tilt_time}, it is evident that the tilt speed remains positive (bar tilt increases) for $\approx 40$ Myr after the collision, and then becomes negative (bar tilt decreases). The tilt speed reversing its sign can be attributed to the influence of restoring forces on the bar from the disk and halo. So, to be in the impulsive regime,  equation (\ref{eq:theta_t}) is integrated from $t = 0$ to $t = 40$ Myr. Here, we consider the collision epoch to be $t = 0$. We use Simpson's rule to perform the integration, using the python package {\em scipy.integrate.simpson}. 

The error in the computation of the bar tilt is dominated by the uncertainty in bar length ($a$). To capture this error, we compute the bar tilt by varying $a$ within one softening length of the simulation. Strictly speaking, the SAM is only valid for constraining the bar tilt in the impulsive regime (upto 40 Myr after the collision). However, the tilt remains roughly constant (with a variation of $\approx 1^\circ$, see Figure \ref{fig:tilt_time}) for at least 100 Myr after the impulsive regime is over. Hence, the tilt predicted by the SAM at the end of the impulsive regime can be used as a reasonable estimate for the tilt at present day. The uncertainty in the bar length is expected to contribute significantly more to the SAM error budget as compared to the variation in the bar tilt after the impulsive regime.

Figure \ref{fig:bar_tilt_model} (left panel) shows the bar tilt calculated from the SAM as a function of $M_{SMC}(< 2 \: \rm{kpc})$. The linear relation between the bar tilt at present day ($\eta$) and $M_{SMC}(< 2 \: \rm{kpc})$ as predicted by the SAM is (blue solid line in Figure \ref{fig:bar_tilt_model}):
\begin{equation}\label{eq:tilt_mass_relation}
    \eta \: [^\circ] = 6.3^\circ\times\frac{M_{SMC}(< 2 \: \rm{kpc}) \: [M_\odot]}{[10^9 \: M_\odot]}
\end{equation}

The bar tilt increases with $M_{SMC}(< 2 \: \rm{kpc})$, which is in accordance with intuitive expectations: a more massive SMC will exert more torque. The sensitive dependence of the bar tilt on $M_{SMC}(< 2 \: \rm{kpc})$ can be used to constrain the mass profile of the SMC. 

The bar tilt in Model 2 present day is $\approx 8.6^\circ$ (Figure \ref{fig:tilt_time}). To reproduce the Model 2 bar tilt, the SAM requires $M_{SMC}(< 2 \: \rm{kpc}) = (1.2 \: - \: 1.6) \times 10^9 \: M_\odot$ within the $3-\sigma$ error interval. The SAM prediction is validated by comparing with the mass profile of the SMC in the Model 2 simulation.

Figure \ref{fig:bar_tilt_model} (right panel) shows the total mass profile (including dark matter, stars and gas) of the SMC in Model 2 at two epochs - the LMC-SMC collision and the present day. The relevant comparison for the SAM prediction is with the SMC mass profile at the time of collision. The collision modifies the total SMC mass profile substantially (a factor of 2). The SAM prediction of $M_{SMC}(< 2 \: \rm{kpc})$ using the Model 2 bar tilt, is indeed consistent at the $3-\sigma$ level with the Model 2 SMC's mass within $\approx 2$ kpc at the LMC-SMC collision epoch (but would be inconsistent with the present day mass profile at the $5\sigma$ level). Thus, the observed LMC bar's tilt can be used to constrain the actual SMC's total mass in its inner regions prior to the collision using the presented SAM. 

Applying the SAM to the observed bar tilt of $5^\circ-15^\circ$ (C18), we find that the SMC's total mass within 2 kpc prior to the LMC-SMC collision must be $M_{SMC}(< 2 \: \rm{kpc}) = (0.8 - 2.4) \times 10^9$ $M_\odot$. To the best of our knowledge, this is the first time a pre-collision dynamical constraint has been obtained for the SMC's total mass profile.

B12 matched the simulated SMC's baryonic (stars $+$ gas) mass profile at present day to the observed constraints (see B12 paper for details). Hence, we can use the simulated SMC's baryonic mass profile at the collision epoch to estimate what the actual SMC's baryonic mass (within 2 kpc) would have been when the collision happened. At the time of the collision, the Model 2 SMC contained a baryonic mass of $\approx 3 \times 10^8 \: M_\odot$ within 2 kpc. Based on the SAM mass requirement to explain the observed bar tilt, prior to the collision, dark matter needs to constitute at least 70\% of the SMC's mass within 2 kpc, requiring the SMC to be a dark matter dominated galaxy.

Note that our SAM makes several simplifying assumptions, and should be treated as a proof-of-concept framework. This SAM was validated by comparison to the results of the B12 simulation, where the impact parameter is known to be 2 kpc. The SAM can be generalized to different impact parameters, now that we have validated it for this specific simulation. This generalization will be the subject of future studies with a larger suite of LMC-SMC collisions. Through this framework, we have shown that it is of paramount importance to accurately constrain the LMC bar's tilt and the LMC-SMC impact parameter to obtain more stringent constraints on the SMC's pre-collision dark matter profile. From Figure \ref{fig:bar_tilt_model}, it is evident that a measurement of the bar tilt that is accurate within $5^\circ$ will improve our SMC mass constraints by a factor of 2.

Constraining the dark matter content of a satellite galaxy like the SMC is consequential for the Lambda Cold Dark Matter ($\Lambda$CDM) cosmological paradigm. The SMC offers a crucial data-point for tightening the constraints at the low mass end of the stellar mass - halo mass - halo concentration relations \citep{Wechsler2002, Diemer2015, Behroozi2019, Wang2021, Wang2024, Bowden2023}. However, despite the SMC being nearby, obtaining dynamical estimates of its mass has been challenging, and only a few constraints exist \citep[e.g.][]{Harris2006, DiTeodoro2019}. Moreover, the present-day mass distribution and kinematics of the SMC has likely been significantly affected by the tidal interactions with the LMC \citep{Subramaniam2009, Zivick2021, Murray2024} and in particular due to the recent collision (as in B12 Model 2, Figure \ref{fig:bar_tilt_model} right panel). In order to place the SMC in context with predictions from the $\Lambda$CDM cosmology, we need constraints on its {\it pre-collision} dark matter profile. Our semi-analytic framework provides a way forward.

Several studies have placed constraints on the present day dynamical mass of the SMC at different radii. \cite{Harris2006} (hereafter HZ06) derived a dynamical mass using a spectroscopic survey of $\approx 2000$ red giant stars in the central 4 kpc $\times$ 2 kpc of the SMC. Using Virial analysis, they derive an enclosed total mass of $1.4 \: - \: 1.9 \times 10^9 \: M_\odot$ within 1.6 kpc of the SMC's center, and an enclosed total mass of $2.7 \: - \: 5.1 \times 10^9 \: M_\odot$ within 3 kpc of the SMC's center. \cite[hereafter DT19]{DiTeodoro2019} used the SMC's HI kinematics and derived a total mass of $1.3 \: - \: 2 \times 10^9 \: M_\odot$ within 4 kpc of the SMC's center. The present day SMC mass profile of the B12 Model 2 simulation is consistent with the predictions of DT19 and within a factor of 2 of HZ06, meaning that the simulated pre-collision mass profile is also likely reasonable.

Assuming the SMC has a total mass of $2 \times 10^9 \: M_\odot$ within 2 kpc, we can make a rough estimate of the dynamical timescale in the SMC's central region:
\begin{equation}
    T_{dyn} \sim 2\pi \sqrt{\frac{R^3}{GM}} = 2\pi \sqrt{\frac{(2 \: \rm{kpc})^3}{G (2 \times 10^9 \: M_\odot)}}
\end{equation}
\noindent which yields $\sim 150$ Myr. The last catastrophic event for the SMC, the LMC-SMC collision, also likely happened $\sim$~150-200 Myr ago. Thus, roughly one dynamical time has passed in the SMC's central region since the collision. This means that the assumption of virial equilibrium is likely not valid, which would affect the accuracy of mass estimates based on stellar and gas kinematics assuming equilibrium dynamics (like HZ06 and DT19). In future work, we plan to quantify the impact of the collision on the SMC's stellar and gas kinematics and its consequent effect on mass inferences at present day.

The assumptions we have made in our SAM essentially rely on the bar acting like a solid body in the impulsive regime. This is reasonable, since the bar in itself is a self-gravitating resonant collection of x1 orbits \citep[e.g.][]{Valluri2016}. Indeed, using proper motions from the VMC survey \citep{Cioni2011}, \cite{Niederhofer2022} found evidence of the LMC bar being supported by x1 orbits. The response of the constituent orbits to the bar offset and tilt over timescales longer than the impulsive regime is an interesting dynamical problem, which we plan to investigate in the future.  

\subsection{Limitations of Our Work and Future Scope} \label{sec:limitations}

As mentioned in section \ref{sec:sims}, the B12 simulations have been very successful in reproducing several observed features of the Clouds. However, as with any simulation, this is just one possible scenario of the LMC-SMC-MW interaction history, and has its limitations. For a detailed list of limitations of the B12 simulations along with their explanations, we refer the reader to \cite{Besla2012, Besla2013}. Here, we discuss some of the caveats that are pertinent to the analysis of the LMC's bar, particularly in the Model 2 simulation.

The low dark matter resolution of the B12 simulations ($10^6$ M$_\odot$ per particle), and the large difference between the dark matter resolution and the stellar resolution (2500 M$_\odot$ per particle) presents limitations that impact the current study of the dynamical structures in the LMC's disk. It has been shown that a low resolution initial condition leads to a higher Poisson noise \citep[e.g.][]{Sellwood2024}, which can cause spurious heating of the disk during the subsequent evolution \citep{Wilkinson2023, Ludlow2023}.

Despite the resolution limitation, we argue that the evolution of the large-scale morphological peculiarities of the bar, like the offset and tilt, can be well studied with the B12 simulations because they arise over short timescales. \cite{Wilkinson2023} find that spurious heating effects are consequential for the long term evolution of a disk over timescales of a few Gyr, which is significantly longer compared to the timescale over which the LMC-SMC collision happens ($\sim$ 50 Myr). Hence, the effect of a SMC collision on a pre-existing LMC bar can be reliably studied with the B12 simulation. Moreover, \cite{Ludlow2023} show that a low resolution dark matter halo does not significantly affect the distribution of gas in the galaxy as well as the Star-formation history. Hence, our inferences about the LMC's gas distribution (section \ref{sec:gas_bar}) are also not expected to be significantly affected by resolution issues. Finally, our conclusions rely on a statistical comparison between Model 2 and Model 1, where we have shown that these unusual morphological characteristics are primarily influenced by the SMC's collision. The low dark matter resolution is not consequential for these morphological peculiarities, since Model 1 also faces the same limitations. Thus, the limited B12 resolution should not affect the main conclusions of our work.

The separation between the Clouds in Model 2 present day is $\approx10$ kpc (Figure \ref{fig:orbit}), whereas the observed separation is $\approx20$ kpc \citep{Kallivayalil2006a, Kallivayalil2006b}. However, we do not expect this detail to significantly affect the scenario we have presented for the LMC's bar, since the dominant torques are applied by the SMC within $\sim 40$ Myr of the collision (as discussed in section \ref{sec:toy_model}), when the SMC is still within 5 kpc of the LMC.

It has been shown that the timing of the LMC-SMC collision may be too recent in the B12 simulation \citep{Choi2022}. In section \ref{sec:offset}, we find that the bar offset in the simulation is significantly larger compared to observations, which is also likely a timing issue. The collision timing may also influence other bar properties like the bar tilt and pattern speed at present day. However, the behavior of the LMC's bar within a certain time interval with respect to the collision epoch can still be well studied with the B12 simulation. Moreover, R25 showed that the Model 2 LMC's present day bar properties (like strength, radius and $m = 2$ Fourier profile) are a good match to observations, re-enforcing the reliability of the B12 simulations for understanding the LMC's bar. Further, comparison of the simulations with the observations can help place a constraint on the timing of the LMC-SMC collision, as we have demonstrated in this work. 

The future evolution of the bar's offset, tilt, pattern speed as well as the response of gas cannot be well studied with the B12 simulations due to the limited number of snapshots beyond the present day. It is beyond the scope of this work to reproduce the exact B12 simulation setup and run the simulation for more time beyond the present day.

An LMC-SMC collision with the assumed impact parameter ($\approx 2$ kpc) is allowed by the present day proper motion error space of the Clouds, but it is not the mean result \citep{Zivick2018}. \cite{Cullinane2022a, Cullinane2022b} built dynamical models for the LMC-SMC-MW interaction history where the LMC was modeled as a rigid potential with stellar tracer particles, and the SMC and MW were modeled purely as rigid potentials. They considered an SMC orbit where the last pericenter about the LMC occurred around 150 Myr ago with a pericentric distance of $\approx8$ kpc. This is consistent with the orbit solution corresponding to the mean proper motions found by \cite{Zivick2018}. In the Cullinane et al. models, the SMC's pericenter resides around 8 kpc below the LMC's disk plane, so this is not a collision scenario. They find that their models are able to qualitatively reproduce the morphology and kinematics of the substructures found in the LMC's periphery in the MagES survey \citep{Cullinane2020}. However, based on our findings for Model 1, it is unlikely that such an orbit can produce a bar tilt consistent with observations or affect the LMC's gas disk significantly to stop the central gas inflows. From our torque analysis (section \ref{sec:toy_model}), we find that the SMC's torques are dominant when it is within 5 kpc of the LMC's center. This scenario is allowed within the 1$\sigma$ proper motion error space, \citep{Zivick2018}. 

Further, most statistical studies of the SMC's orbit have assumed a static potential for the LMC \citep[e.g.][]{Kallivayalil2006a, Kallivayalil2013, Zivick2018, Patel2020, Cullinane2022a, Cullinane2022b}. The SMC's motion around the LMC likely induces significant dynamical friction wakes on the latter's halo, and these wakes can back-react on the SMC itself, which may not be fully captured by the Chandrasekhar dynamical friction formula \citep{Chandrasekhar1943}. Such wakes can significantly change the likelihoods of possible SMC orbits about the LMC and impact parameters. Indeed, for the LMC-MW system it has been shown that the back-reaction of the MW halo perturbations on the LMC can change the LMC's orbit \cite{GC2021}. The effect is expected to be stronger for the LMC-SMC system given their multiple orbits about each other and smaller impact parameter, as compared to the LMC-MW system, which is a likely first infall \citep[e.g.][]{Besla2007}.

The wakes in the LMC's dark matter halo can also apply significant torques on the LMC's bar and disk. We have not taken this effect into account while developing the SAM for bar tilt (section \ref{sec:toy_model}). However, the dark matter particle resolution in the B12 simulations ($\approx 10^6 \: M_\odot$) is not sufficient for resolving these dynamical friction wakes. Hence modeling the gravitational force from the wakes using B12 simulations is currently not feasible.

A higher resolution simulation of the LMC-SMC-MW interaction history (having a recent, direct collision between the clouds) which is significantly more fine-tuned to observations and has sufficient snapshots beyond the present day is required to study the formation and evolution of the LMC's bar in depth. This is work in progress. Further, in this future work (Rathore et al. 2025(c), \textit{in prep}), we shall include detailed resolution convergence studies to ascertain that the long term evolution (over a few Gyr) of the LMC bar properties are not affected by the particle resolution.

Some models of the KRATOS suite of simulations that do not include the Milky Way also have close encounters between the LMC and SMC, providing a promising opportunity to study the evolution of the LMC's bar \citep[e.g.][]{Arranz2025b}. However, the KRATOS simulations do not include gas. Hydrodynamical processes can keep the disk kinematically colder by radiatively cooling the gas and forming new stars, which can affect the evolution of the bar.

\section{Conclusion} \label{sec:conclusion}

The LMC possesses an unusual bar. The bar is offset from the center of the outer disk isophotes by $\sim 1$ kpc, is tilted out of the disk plane by $5^\circ - 15^\circ$, and has no signature in the spatial distribution of neutral gas. The LMC's bar pattern speed has been found to be unusually slow, or maybe even counter-rotating. Using numerical simulations, we investigate the hypothesis that a recent ($\sim$ 100 Myr ago) collision (impact parameter $\approx$ 2 kpc) between the LMC and SMC is the primary driver of the LMC bar's unusual properties.

We utilize hydrodynamic simulations of the LMC-SMC-MW interaction history by B12. The LMC and SMC are modeled with live exponential stellar disks, Smoothed-Particle-Hydrodynamic gas disks (with star formation and radiative cooling) and live dark matter halos. The total masses of the LMC and SMC in the simulation are $1.8 \times 10^{11}$ $M_\odot$ and $2 \times 10^{10}$ $M_\odot$ respectively. The Milky Way (MW) is modeled as a static NFW potential of mass $1.5 \times 10^{12}$ $M_\odot$. B12 present two scenarios: Model 1 and Model 2. In Model 1, the LMC and SMC remain far from each other, with their closest separation being $>$ 20 kpc. In Model 2, the LMC and SMC undergo a recent collision (impact parameter $\approx$ 2 kpc) around 100 Myr ago. The SMC's orbit about the LMC in Model 2 is significantly inclined ($\approx$ 50$^\circ$) with respect to the LMC's disk plane prior to the collision, which means the SMC can affect both in-plane as well as vertical motions of the LMC's bar. In Model 1, given the large distance between the LMC and SMC, we do not expect the SMC to significantly affect the LMC's bar.

In both Model 1 and 2, the Clouds are on first infall, making their first inwards crossing of the MW's virial radius 1 Gyr ago. Further, in both models the LMC disk forms a stellar bar $>4$ Gyr ago owing to secular evolution, well before the Clouds infall to the MW. The only difference between these scenarios is the absence of a direct collision in Model 1. The B12 LMC Model 2 has been shown to reasonably represent the structure \citep{Besla2016}, kinematics \citep{Choi2022}, and bar properties \citep{Rathore2025} of the observed LMC, making it appropriate for this study. 

In this paper we analyze and compare the properties of the LMC's bar in both B12 Model 1 and Model 2 to discern the impact of a collision on the origin of the bar's unusual nature. We summarize our results as follows:

\begin{itemize}
    \item \textit{Post SMC collision, the LMC's bar develops a large offset:} We fit isodensity ellipses to the simulated LMC's stellar mass distribution (Figure \ref{fig:ellip_offset}). We define the bar offset as the separation between the centers of the isodensity ellipses corresponding to the bar and the outer disk. The bar offset increases from a mean value of $\approx$ 0.3 kpc to $\approx$ 1.5 kpc as the SMC collides with the LMC (Figure \ref{fig:offset_time}). In Model 1, where the SMC does not collide with the LMC, the bar offset remains small ($0.30 \pm 0.16$ kpc), which is inconsistent with observations.
    \item \textit{The LMC bar's offset constrains the LMC-SMC collision to have happened 150-200 Myr ago:}
    Comparing the Model 2 simulated bar offset ($\approx$ 1.5 kpc), to the observed \citep[0.76 kpc,][]{Rathore2025}, we infer that the LMC's disk needs to evolve for at least $50$ Myr after the simulation present day for the offset to reduce to the observed value, which suggests the epoch of the LMC-SMC collision to be 150-200 Myr ago (Figure \ref{fig:offset_time}). This is the same time range as that concluded by \cite{Choi2022} based on the internal stellar kinematics of the LMC.
    \item \textit{Post SMC collision, the LMC's bar and outer disk center separate from the DM halo center:} We infer the simulated LMC's DM halo center through an iterative shrinking sphere process. Post collision, the bar center and the DM halo center as well as the outer disk center and the DM halo center are separated by $\approx 1$ kpc (Figure \ref{fig:bar_dm_offset}). The bar-DM separation evolves with time, and negligible separation is expected 150-200 Myr after the collision. However, the outer disk - DM separation persists and is significant ($\approx$ 1 kpc) even 150-200 Myr after the collision.
    \item \textit{Post SMC collision, the LMC's bar develops a large tilt:}
    We fit isodensity ellipses to the simulated LMC's edge-on stellar mass distribution (Figure \ref{fig:ellipse_tilt}), and find that the bar tilt increases from a mean value of $1^\circ \pm 1^\circ$ to $8.6^\circ \pm 1^\circ$ as the SMC collides with the LMC (Figure \ref{fig:tilt_time}), which is consistent with observations (5$^\circ$ - 15$^\circ$, C18). In Model 1 (no collision), the bar tilt remains small (within 2$^\circ$). 
    \item \textit{Post SMC collision, the LMC is well-described by a Tilted-Ring Morphology:} We construct the angular momentum profile of the Model 2 LMC's disk at present day (Figure \ref{fig:briggs}). The angular momentum vector of the LMC's inner disk containing the bar (R $<$ 2 kpc) is misaligned with respect to the outer disk at 4 - 6 kpc by $\approx 10^\circ$. The angular momentum vector of the transition region (2 - 4 kpc) between the inner disk and outer disk is misaligned with respect to the outer disk by $\approx 4^\circ$. Whereas, the angular momentum vector of the outer disk is consistent with the average disk plane. Thus, the angular momentum profile of the simulated LMC is consistent with the observed LMC's \enquote{tilted-ring morphology} (JA25), which is naturally produced by the LMC-SMC collision.
    \item \textit{The pattern speed of the LMC's bar is affected by an SMC collision}: We quantify the pattern speed of the LMC's bar using the time evolution of the $m = 2$ phase angle (Figure \ref{fig:ps_model2}). The SMC collision significantly affects the bar pattern speed, whereas in Model 1 (no collision) the bar pattern speed is unchanged. In the Model 2 configuration, the pattern speed reduces by a factor of 2 at the present day. 
\end{itemize}

We conclude that a recent collision with the SMC can explain the aforementioned unusual properties of the LMC's bar, whereas a scenario where the SMC's impact parameter remains large ($> 5$ kpc), results in LMC bar properties that are inconsistent with observations. We also provide a framework (Figure \ref{fig:gas_maps} and \ref{fig:gas_time}) to quantitatively compare the LMC's gas distribution in different LMC-SMC interaction scenarios. This framework can be generalized to more advanced hydrodynamic simulations.

We developed a semi-analytic model (SAM) based on the impulse approximation to study the torques applied by the SMC on the LMC's bar. The SAM provides a useful framework to understand the changes to the LMC's bar during an LMC-SMC collision. Using the SAM, we find:  

\begin{itemize}
    \item \textit{The LMC bar's tilt is sensitive to the SMC's pre-collision mass within 2 kpc:} Given the Model 2 tilt of $8.6^\circ$, the SAM predicts $M_{SMC}(< 2 \: \rm{kpc}) = (1.2 \: - \: 1.6) \times 10^9 \: M_\odot$, which is consistent with the simulated SMC's total mass profile prior to the collision. The validation of the SAM means this framework can be applied to observations to constrain the actual SMC's total mass profile prior to the collision. Using the observed tilt of $5^\circ-15^\circ$, the SAM predicts $M_{SMC}(< 2 \: \rm{kpc}) = (0.8 \: - \: 2.4) \times 10^9 \: M_\odot$ (Figure \ref{fig:bar_tilt_model}) for an LMC-SMC impact parameter of 2 kpc. Given that the SMC's expected baryonic mass within 2 kpc is unlikely to be greater than $3 \times 10^8$ M$_\odot$, even the minimum bar tilt of $5^\circ$ confirms that the SMC is a dark matter dominated galaxy.
    \item \textit{The SMC and LMC bar are in a state of dis-equilibrium:} The dynamical timescale of the inner SMC is similar to the time elapsed since the collision (150 - 200 Myr). As such, insufficient time has elapsed for the system to achieve Virial equilibrium. The SMC's mass profile has evolved considerably since the collision, becoming shallower by a factor of two in the inner regions. Thus, equilibrium dynamics cannot be applied to describe the SMC's structure, kinematics and mass distribution. Further, the LMC's bar is still evolving. The bar has likely not even completed one rotation since the collision. The bar's pattern speed, offset and the strength of its gaseous counterpart are changing faster than the time period of bar rotation. This means equilibrium dynamics cannot be applied to the LMC's disk, complicating efforts to interpret e.g. the bar pattern speed from observations.
\end{itemize}

This work highlights the critical need for observational efforts to better constrain the tilt of the LMC's bar and the LMC-SMC impact parameter. If the tilt is measured accurately (with an error of at most $5^\circ$), the SMC's pre-collision mass profile could be constrained within a factor of 2. This would enable us to place the SMC in context within the cosmological stellar mass - halo mass - halo concentration relations.

Future work involves building higher resolution simulations of the Clouds' interaction history with more advanced hydrodynamic prescriptions, enabling a better match to the observational constraints. This would establish the Clouds as a precision laboratory for galactic bar and disk dynamics, and the prototype for understanding the role of galactic collisions in galaxy evolution.

\begin{acknowledgements}
Himansh Rathore would like to thank Jerry Sellwood, Yumi Choi, Knut Olsen, Dennis Zaritsky, Tod Lauer, Annapurni Subramaniam, Smitha Subramanian, Ann-Marie Madigan, {\'O}. Jim{\'e}nez-Arranz, Martin Weinberg, Monica Valluri, Elena D'Onghia, Mathieu Renzo, Mariarosa Cioni and Amy Smock for interesting discussions on the LMC's bar. We would like to thank the anonymous referee for providing insightful comments that improved the quality as well as clarity of the paper. Himansh Rathore and Gurtina Besla are supported by NASA FINESST 80NSSC24K1469, NASA ATP 80NSSC24K1225 and NSF CAREER AST 1941096. This work utilized the Puma and ElGato High Performance Computing clusters at the University of Arizona. We respectfully acknowledge the University of Arizona is on the land and territories of Indigenous peoples. Today, Arizona is home to 22 federally recognized tribes, with Tucson being home to the O’odham and the Yaqui. The University strives to build sustainable relationships with sovereign Native Nations and Indigenous communities through education offerings, partnerships, and community service.

\software{This work made use of python, and its packages like numpy \citep{vanderWalt2011, Harris2020}, scipy \citep{Virtanen2020} and matplotlib \citep{Hunter2007}.}

\end{acknowledgements}

\bibliography{references}{}
\bibliographystyle{aasjournal}

\end{document}